\documentclass[a4paper,11pt]{article}
\usepackage{jcappub}
\pdfoutput=1
\usepackage{enumerate}
\usepackage{graphicx}
\usepackage{slashed}
\usepackage{float}
\usepackage[dvipsnames]{xcolor}
\usepackage[normalem]{ulem}
\usepackage{subfigure,orcidlink}
\usepackage{multirow,array}
\usepackage{hyperref}
\usepackage{tabularray}
\UseTblrLibrary{booktabs}

\title{Singlet-doublet dark matter induced radiative neutrino mass and TeV scale leptogenesis}

\author[]{Partha Kumar Paul$^{\orcidlink{0000-0002-9107-5635}}$,}
\author[]{Narendra Sahu$^{\orcidlink{0000-0002-9675-0484}}$}
\author[]{and Shashwat Sharma$^{\orcidlink{0009-0002-2266-0467}}$}
\affiliation[]{Department of Physics, Indian Institute of Technology Hyderabad, Kandi, Sangareddy, Telangana-502285, India.}
\emailAdd{ph22resch11012@iith.ac.in}
\emailAdd{nsahu@phy.iith.ac.in}
\emailAdd{ph23resch11016@iith.ac.in}
     
\abstract{The singlet–doublet dark matter (SDDM) model is a well-motivated WIMP framework that accommodates viable DM over a broad range of parameter space. In this work, we explore the possibility of TeV-scale leptogenesis within two realizations of the SDDM setup: Majorana SDDM scenario and Dirac SDDM scenario. The light neutrino mass, in either case, arises radiatively at one loop level. The particles running in the loop are responsible for DM relic and TeV-scale leptogenesis while satisfying other phenomenological constraints. In the Majorana setup, the Standard Model is extended by three generations of singlet fermions $N_i$ and doublet fermions $\Psi_i$, and a singlet scalar $\phi$. The \textit{CP}-violating, out-of-equilibrium decays of the heavier singlets ($N_{2,3}$) generate baryon asymmetry via the leptogenesis route, while the first generation of singlet-doublet fermions give rise to the usual SD Majorana DM. In the Dirac setup, the standard model is extended by three generations of complex scalars ($\phi_i$) and right-handed Dirac partners ($\nu_{R_i}$) of SM neutrinos ($\nu_{L_i}$), along with a pair of singlet–doublet fermions $\chi$ and $\Psi$. The \textit{CP}-violating out-of-equilibrium decays of the scalar fields $\phi_i$ generate baryon asymmetry via the Dirac leptogenesis route. We show that in the Majorana setup, successful leptogenesis is possible even in the sub-TeV regime, whereas in the Dirac setup, the leptogenesis scale is at a few TeV. With the particle mass at the TeV scale, the model remains promising for collider experiments, particularly through signatures such as prompt decays and displaced vertex searches. In addition, the presence of Dirac neutrinos can contribute to $\Delta N_{\rm eff}$, providing complementary cosmological signatures.}

\begin{document}
\maketitle
\flushbottom
\section{Introduction}
\label{sec:intro}

The Standard Model (SM) of particle physics successfully describes the masses of elementary particles and their interactions, but it fails to account for dark matter (DM) which is about $26.8\%$ \cite{Planck:2018vyg} of the energy budget of the universe, the origin of neutrino masses as implied by the oscillation data \cite{deSalas:2017kay,T2K:2015sqm,IceCube:2015fuw,Bahcall:2004mz,KamLAND:2002uet,Super-Kamiokande:2001bfk}, and the observed matter–antimatter asymmetry of the Universe as revealed by BBN and satellite-borne experiment Planck \cite{Planck:2018vyg}. Leptogenesis \cite{Sakharov:1967dj,Fukugita:1986hr,PhysRevD.45.455,PhysRevD.46.5331,Flanz:1994yx,Davidson:2008bu,Buchmuller:2004nz, Barbieri:1999ma,Pilaftsis:2003gt,Pilaftsis:2005rv} provides an attractive dynamical mechanism in which a lepton asymmetry is produced in the early Universe and later converted into a baryon asymmetry by electroweak sphaleron processes \cite{Kuzmin:1985mm}. Seesaw models (type-I \cite{Minkowski:1977sc,Schechter:1980gr,Mohapatra:1980yp} , type-II \cite{Mohapatra:1980yp,Lazarides:1980nt,Wetterich:1981bx,Schechter:1981cv,Ma:1998dx}, type-III \cite{Foot:1988aq}) are the most economical extensions in the beyond standard model framework to explain baryon asymmetry simultaneously via the leptogenesis route and the non-zero neutrino masses inferred from the oscillation data. However, in the canonical seesaw models, generating viable DM typically requires additional model building beyond the minimal framework\footnote{Notable exceptions exist, such as the $\nu$MSM \cite{Asaka:2005pn, Shaposhnikov:2008pf}, where the lightest right-handed neutrino with a keV-scale mass serves as a warm DM candidate.}. Nevertheless, in the canonical seesaw models, the scale of leptogenesis is very high in order to give rise to the sub-eV masses of light neutrinos. An alternative is to consider scotogenic models see e.g. \cite{Ma:2006km,Farzan:2012sa,Ma:2013yga,Ma:2016mwh,Barbieri:2006dq} and references therein, where the neutrino mass arises radiatively at one-loop level \cite{Krauss:2002px,Kubo:2006rm,Cheung:2004xm,Gustafsson:2012vj}. In these models, the particles running in the loop are decoupled from the SM due to an imposed symmetry. As a result, the lightest particle running in the loop is the natural candidate for DM. In such models, decay of the heavier particles running in the neutrino loop may give rise to lepton asymmetry, thus providing a common platform for neutrino mass, DM, and baryon asymmetry of the universe, see e.g. \cite{Hugle:2018qbw,Borah:2018rca,JyotiDas:2021shi,Borah:2016zbd,Baumholzer:2018sfb,Sarma:2020msa,Borah:2021qmi,Singh:2023eye,Sahu:2008aw,Narendra:2018vfw,Borah:2018smz,Tapender:2026ets} and references therein. For other low scale leptogenesis mechanisms see \textit{e.g.}, incorporating flavor effects \cite{Blanchet:2008pw}, ARS
leptogenesis \cite{Akhmedov:1998qx,Asaka:2005pn,Hernandez:2016kel}, by adding extra scalar fields \cite{Clarke:2015hta,Hugle:2018qbw,Borah:2021qmi,Vatsyayan:2022rth}, resonant leptogenesis \cite{Pilaftsis:2003gt}, right-handed sector leptogenesis \cite{Frigerio:2006gx,McDonald:2007ka}, wash-in leptogenesis \cite{Domcke:2020quw}, and by decoupling neutrino mass and leptogenesis \cite{Ma:2006te,Ma:2006ci,Hirsch:2006ft,Hambye:2006zn,Paul:2024iie}.

In this paper, we consider a variant of scotogenic neutrino mass model which is inspired by singlet-doublet fermion dark matter (SDDM) \cite{Bhattacharya:2018fus,Cynolter:2015sua,Bhattacharya:2015qpa,Bhattacharya:2017sml,Bhattacharya:2018cgx,Bhattacharya:2016rqj,Dutta:2020xwn,Borah:2021khc,Borah:2021rbx,Borah:2022zim,Borah:2023dhk,Paul:2024iie,DEramo:2007anh,Cohen:2011ec,Freitas:2015hsa,Calibbi:2015nha,Cheung:2013dua,Banerjee:2016hsk,DuttaBanik:2018emv,Horiuchi:2016tqw,Restrepo:2015ura,Abe:2017glm,Konar:2020wvl,Konar:2020vuu,Calibbi:2018fqf,Ghosh:2021wrk,Kundu:2024nig,Bhattacharya:2021ltd,Enberg:2007rp,Oncala:2021tkz,Paul:2024prs,Paul:2025spm,Dey:2025pcs} in which the dark sector contains a fermionic singlet $N$ and an electroweak doublet $\Psi$ that mix after electroweak symmetry breaking. This setup is characterized by just three main parameters: the DM mass, the singlet–doublet mixing angle, and the mass splitting between the DM and the doublet. The model yields viable DM candidates across a wide mass range from the GeV to the TeV scale. The minimal SDDM framework is further augmented by adding a singlet scalar $\phi$, allowing neutrino masses to be generated radiatively at one loop. This additional scalar not only enriches the neutrino sector but also opens the door for leptogenesis within the same setup. We further show that TeV scale leptogenesis can be achieved in scenarios where the DM particle is either:
\begin{itemize}
    \item Majorana (singlet-doublet Majorana DM).
    \item Dirac (singlet-doublet Dirac DM).
\end{itemize}

In the first scenario, in which the SDDM is Majorana in nature, the model constitutes three generations of heavy singlet–doublet fermion pairs ($N_i$ and $\Psi_i$) along with a singlet scalar ($\phi$) \cite{Borah:2022zim}. We impose a $\mathcal{Z}_2$  symmetry under which these additional particles are odd while all the SM particles are even. As a result, the light neutrino masses arise radiatively at the one-loop level. In this setup, the particles running in the neutrino loop are responsible for the DM and low-scale leptogenesis while satisfying constraints from $(g-2)_{\mu}$ and charged lepton flavor violation ($\mu\rightarrow e\gamma$) \cite{Lindner:2016bgg}. The lightest singlet-doublet fermion ($N_1, \Psi_1$) pair constitutes the DM content of the universe while the \textit{CP}-violating out-of-equilibrium decays of heavier singlets ($N_{2,3}$) produces an asymmetry in $\Psi$ through the decay channel $N\rightarrow\Psi H$, which is subsequently transferred to the SM leptons\footnote{Since $\Psi$s are vector-like fermions, they do not participate in sphaleron processes \cite{Arina:2012aj}. Thus, $\Psi$ asymmetry can not be directly converted to the baryon asymmetry. It is first transferred to the lepton asymmetry, and then the lepton asymmetry is converted to the baryon asymmetry via the electroweak sphalerons.} and converted to baryon asymmetry via the EW sphaleron process\cite{Kuzmin:1985mm}. We show that successful leptogenesis is possible even in the sub-TeV region, yet correctly produces the DM relic.  

In the second scenario, in which the SDDM is Dirac in nature, the model constitutes a pair of singlet-doublet Dirac fermions ($\chi,\Psi$), three generations of right-handed Dirac partners ($\nu_{R_i}$) of SM neutrinos ($\nu_{L_i}$) together with three generations of singlet scalars ($\phi_i$) \cite{Borah:2023dhk}. We impose a $\mathcal{Z}_4$ symmetry under which these particles carry non-trivial charges. As a result, neutrino mass at tree level is forbidden and arises radiatively at one loop. In this setup, a mixture of $\chi$ and the neutral component of $\Psi$ constitutes the DM relic. Because the fermions are Dirac particles, the total lepton number remains conserved. The CP-violating out-of-equilibrium decay of the lightest singlet scalar $\phi_1$ produces equal and opposite asymmetries in the left and right-handed lepton sectors. While the total lepton number vanishes, the asymmetry stored in the left-handed sector is partially converted into a baryon asymmetry via electroweak sphaleron transitions, whereas the right-handed sector remains inert with respect to sphaleron processes. We demonstrate that successful leptogenesis can occur at a scale of a few TeV, while still producing the correct DM relic density.

The paper is organized as follows. In Sec.~\ref{sec:modelA_majorana}, we discuss the singlet–doublet Majorana DM setup. The model is introduced in Sec.~\ref{subsec:modelA}, followed by a discussion of neutrino mass in Sec.~\ref{subsec:neutrinoA}. The muon $(g-2)$ and charged lepton flavor violation (cLFV) for the Majorana case are presented in Secs.~\ref{subsec:muong-2A} and \ref{subsec:lfvMajorana}, respectively. The DM phenomenology of the singlet–doublet Majorana scenario is discussed in Sec.~\ref{subsec:DM_A}. Leptogenesis from $N_{2,3}$ decay is studied in Sec.~\ref{sec:majoranalepto}. We then turn to the singlet–doublet Dirac DM model in Sec.~\ref{sec:model_B}, with details of the model presented in Sec.~\ref{subsec:model_B}. The generation of neutrino mass in this case is discussed in Sec.~\ref{subsec:diracnumass}. The $(g-2)_\mu$ and cLFV are analyzed in Secs.~\ref{subsec:muong-2Dirac} and \ref{subsec:lfvDirac}, respectively. The corresponding DM phenomenology is discussed in Sec.~\ref{subsec:DM_B}. Dirac leptogenesis from $\phi_i$ decay is presented in Sec.~\ref{sec:leptoD}. Section \ref{sec:coll} discusses the collider signatures of the models. Finally, we conclude in Sec.~\ref{sec:concl}.

\section{Model A: singlet-doublet Majorana dark matter and leptogenesis}\label{sec:modelA_majorana}
\subsection{The model}\label{subsec:modelA}

We extend the Standard Model by 3 generations of right-handed neutrinos, $N_i$, 3 generations of doublet fermions, $\Psi_i$, $i=1,2,3$, and a singlet scalar ($\phi$). We impose a discrete symmetry $\mathcal{Z}_2$ under which the $N_i$'s, $\Psi_i$'s, and $\phi$ are odd, while all other particles are even. We assume a mass hierarchy among different generations of singlet-doublet fermions. $N_1$ and $\Psi_1$ constitutes the lightest singlet-doublet pair while the other pairs $N_{2},\Psi_2$ and $N_{3},\Psi_3$ remain heavy. The neutral component of the fermion doublet $\Psi_1=(\psi_1^0 \quad\psi_1^-)^T$ mixes with the singlet $N_1$ to form the singlet-doublet DM. 
\begin{table}[h!]
		\small
		\begin{center}
			\begin{tabular}{||@{\hspace{0cm}}c@{\hspace{0cm}}|@{\hspace{0cm}}c@{\hspace{0cm}}|@{\hspace{0cm}}c@{\hspace{0cm}}|@{\hspace{0cm}}c@{\hspace{0cm}}||}
				\hline
				\hline
				\begin{tabular}{c}
                {\bf ~~~~Symmetry~~~~}\\
					{\bf ~~~~Group~~~~}\\ 
					\hline
					
					$SU(2)_{L}$\\ 
					\hline
					$U(1)_{Y}$\\ 
					\hline
					$\mathcal{Z}_2$\\ 
				\end{tabular}
				&
				&
				\begin{tabular}{c|c|c}
					\multicolumn{3}{c}{\bf Fermion Fields}\\
					\hline
					~~~$L$~~~& ~~~$\Psi_i =(\psi^0_i~~\psi^-_i)^T $~~~ & ~~~$N_i$~~~ \\
					\hline
					$2$&$2$&$1$\\
					\hline
					$-1$&$-1$&$0$\\
					\hline
					$+$&$-$&$-$\\
				\end{tabular}
				&
				\begin{tabular}{c|c|c}
					\multicolumn{2}{c}{\bf Scalar Field}\\
					\hline
					~~~$H$~~~& ~~~$\phi$~~~\\
					\hline
					$2$&$1$\\
					\hline
					$1$&$0$\\
                    \hline
					$+$&$-$\\
				\end{tabular}\\
				\hline
				\hline
			\end{tabular}
			\caption{Particles and their charge assignments under the group $SU(2)_L\otimes U(1)_Y\otimes \mathcal{Z}_2$.}
			\label{tab:tab1}
		\end{center}    
	\end{table}
The particles and their charges under $SU(2)_L\otimes U(1)_Y\otimes\mathcal{Z}_2$ are given in table~\ref{tab:tab1}. The relevant Lagrangian describing the interactions between the particles is given as
\begin{eqnarray}
\mathcal{L}=\iota\bar{\Psi}\gamma^\mu D_\mu\Psi + \iota\bar{N}\gamma^\mu\partial_\mu N -M_{\Psi}\bar{\Psi}\Psi-\frac{1}{2}M_{N}\overline{N^C}N-y_{\alpha\beta}\bar{\Psi}_{\alpha}\tilde{H}N_{\beta}-\lambda_{i\alpha}\bar{L}_{i}\Psi_\alpha\phi + {\rm H.c.},\nonumber\\ \label{eq:lagrangian}
\end{eqnarray}
Here, we have suppressed the generation indices.\footnote{Without the loss of generality, we can consider bare mass matrix for $\Psi$ and $N$ to be diagonal, i.e., $M_{\Psi_{ii}}\bar{\Psi}_i\Psi_i$ and $M_{N_{ii}}\bar{N}_i^cN_i$.} After the electroweak phase transition, $N_1$ mixes with the neutral component of $\Psi_1$ to give rise three mass eigenstates: $\chi_1,\,\chi_2$ and $\chi_3$ with masses $M_{\chi_1},\,M_{\chi_2}$ and $M_{\chi_3}$ respectively. The lightest of these mass eigenstates $\chi_3$ becomes the DM. The details can be found in appendix~\ref{app:majo}. The \textit{CP}-violating-out of equilibrium decays of heavier singlets $N_{2,3}$ to $\Psi_{2,3}$ via the decay channel $N\rightarrow\Psi H$, followed by $\Psi\rightarrow L\phi$ produces the lepton asymmetry. This lepton asymmetry is transferred to the baryon asymmetry via the EW sphaleron process. The imposed $\mathcal{Z}_2$ symmetry plays crucial role in the model. Firstly, it forbids the generation of neutrino masses at tree level, allowing them to arise radiatively at one-loop level, as will be discussed in the following section. Secondly, it stabilizes the singlet-doublet DM candidate. In the absence of the $\mathcal{Z}_2$, neutrino masses would be generated at tree level via the type-I seesaw mechanism, pushing the new physics scale to $\mathcal{O}(10^{12})$ GeV without fine-tuning of the Yukawa couplings, rendering the model inaccessible to current and future experiments. On the other hand, the radiative generation of neutrino masses naturally suppresses them without fine-tuning, and simultaneously keeps the dark sector particles at the TeV scale, making the model testable at colliders and direct detection experiments.

\subsection{Neutrino mass}\label{subsec:neutrinoA}

\begin{figure}[H]
    \centering
    \includegraphics[scale=1.0]{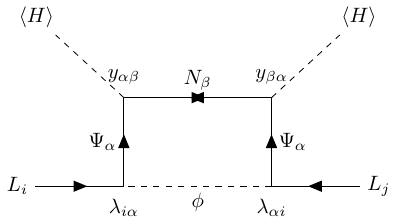}
    \caption{Majorana neutrino mass at one loop.}
    \label{fig:numass}
\end{figure}

Since the scalar $\phi$ doesn't acquire a VEV, neutrino mass at tree level is not possible, but can be produced at one-loop level as shown in figure~\ref{fig:numass} \cite{Fraser:2014yha,Konar:2020wvl,Borah:2022zim}.
From figure~\ref{fig:numass}, the Majorana mass of the light neutrino can be given as:
\begin{eqnarray}
\label{numasseq}
\left(m_{\nu}\right)_{ij}&=&\sum_{\alpha,\beta=1}^3\frac{\lambda^T_{i \alpha }y^T_{\alpha\beta}y_{\beta\alpha}\lambda_{\alpha j}v^2}{4\pi^2}\left(-\frac{2 M_{\Psi_{\alpha} }^2 }{\left(M_{\Psi_{\alpha} }-M_{N_{\beta} }\right) \left(M_{\Psi_{\alpha} }^2-M_{\phi }^2\right)}-\frac{M_{N_{\beta} }^3 \log \left(\frac{M_{N_{\beta} }^2}{M_{\Psi_{\alpha} }^2}\right)}{\left(M_{N_{\beta} }-M_{\Psi_{\alpha} }\right){}^2 \left(M_{N_{\beta} }^2-M_{\phi }^2\right)}\right.\nonumber\\&&-\left.\frac{ M_{\phi }^2\left(M_{N_{\beta} } M_{\Psi_{\alpha} }^2+M_{N_{\beta} } M_{\phi }^2+2 M_{\Psi_{\alpha} } M_{\phi }^2\right) \log \left(\frac{M_{\phi }^2}{M_{\Psi_{\alpha} }^2}\right)}{\left(M_{\phi }^2-M_{N_{\beta} }^2\right) \left(M_{\phi }^2-M_{\Psi_{\alpha} }^2\right){}^2}\right)
\end{eqnarray}
Using Eq.~\ref{numasseq}, we can write the neutrino mass matrix as:
\begin{eqnarray}
    \label{eq:numassshort}
    m_{\nu_{ij}}=(\lambda_{\alpha i})^T\Lambda_{\alpha\alpha}\lambda_{\alpha j}
\end{eqnarray}
where the $\Lambda$ is defined as:
\begin{eqnarray}
    \label{numassfactor}
    \Lambda_{ii}&=&
        \sum_{k=1}^3\frac{y_{ik} y_{ki}v^2}{4\pi^2}\left(-\frac{2 M_{\Psi_i }^2 }{\left(M_{\Psi_i }-M_{\chi_k }\right) \left(M_{\Psi_i }^2-M_{\phi }^2\right)}-\frac{M_{\chi_k }^3 \log \left(\frac{M_{\chi_k }^2}{M_{\Psi_i }^2}\right)}{\left(M_{\chi_k }-M_{\Psi_i }\right){}^2 \left(M_{\chi_k }^2-M_{\phi }^2\right)}\right.\nonumber\\&&-\left.\frac{ M_{\phi }^2\left(M_{\chi_k } M_{\Psi_i }^2+M_{\chi_k } M_{\phi }^2+2 M_{\Psi_i } M_{\phi }^2\right) \log \left(\frac{M_{\phi }^2}{M_{\Psi_i }^2}\right)}{\left(M_{\phi }^2-M_{\chi_k }^2\right) \left(M_{\phi }^2-M_{\Psi_i }^2\right){}^2}\right)
\end{eqnarray}
Using the Casas-Ibarra (C.I.) parameterization \cite{Casas:2001sr}, we define the 
coupling matrix $\lambda_{i\alpha}$ as:
\begin{eqnarray}
    \label{CIparameterization}
    \lambda = \sqrt{\Lambda}^{-1}R D_{\sqrt{m}} U_{\rm PMNS}^\dagger
\end{eqnarray}
where $R$ is a complex rotation matrix, given as:
\begin{eqnarray}
        R &=&
    \begin{pmatrix}
        \cos z_1 & -\sin z_1 & 0 \\
        \sin z_1 & \cos z_1 & 0\\
        0 & 0 & 1
    \end{pmatrix}
    \begin{pmatrix}
        \cos z_2 & 0 & \sin z_2 \\
        0 & 1 & 0\\
        -\sin z_2  & 0  & \cos z_2 
    \end{pmatrix}
    \begin{pmatrix}
        1 & 0 & 0\\
        0 & \cos z_3  & -\sin z_3 \\
        0 & \sin z_3  & \cos \theta_3
        \end{pmatrix}\label{eq:Rmatrix}
\end{eqnarray}
where, $z_i=\alpha_i+\iota\beta_i$, $D_{\sqrt{m}}={\rm diag}(\sqrt{m_1},\sqrt{m_2},\sqrt{m_3})$, where $m_1,m_2,m_3$ are the light neutrino masses, $U_{\rm PMNS}$ is the PMNS matrix defined as:
{{\begin{eqnarray} 
U_{\rm PMNS} &=& 
    \begin{pmatrix}
        c_{12} c_{13} & s_{12} c_{13} & s_{13} e^{-\iota \delta} \\ -s_{12} c_{23} - c_{12} s_{13} s_{23} e^{\iota \delta} & c_{12} c_{23} - s_{12} s_{13} s_{23} e^{\iota \delta} & c_{13} s_{23} \\ s_{12}s_{23} - c_{12} s_{13} c_{23} e^{\iota \delta} & -c_{12} s_{23} - s_{12} s_{13} c_{23} e^{\iota \delta} & c_{13} c_{23}
    \end{pmatrix}
    \begin{pmatrix}
        e^{\iota \eta_1} & 0 & 0 \\ 0 & e^{\iota \eta_2} & 0 \\ 0 & 0 & 1
    \end{pmatrix},
\end{eqnarray}}}
where $c_{ij} = \cos \theta_{ij}$ and $s_{ij} = \sin \theta_{ij}$ and $\delta$ is the Dirac $CP$ phase and $\eta_{1,2}$ are the Majorana $CP$ phases. In our analysis, we have used the best-fit value for the $U_{\rm PMNS}$ matrix, where $s^2_{12}=0.308$, $s^2_{23}=0.470$, $s^2_{13}= 0.02215$, $\delta/{\pi}= 1.178$ and the Majorana phases $\eta_i=0~\forall~ i=1,2$ \cite{Esteban:2024eli}.

\subsection{Muon anomalous magnetic moment}
\label{subsec:muong-2A}

\begin{figure}[h]
    \centering
    \includegraphics[scale=1.0]{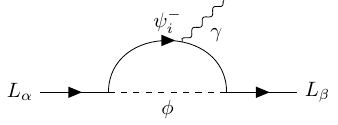}
    \caption{The Feynman diagram giving rise $(g-2)_\mu$ and charged lepton flavor violation.}
    \label{fig:lfvmajorana}
\end{figure}

In our setup, the new positive contribution to the muon $(g-2)$ arises from the one-loop diagram involving the charged doublet fermions $\psi^{-}_i$ and the singlet scalar $\phi$ running in the loop, as shown in figure~\ref{fig:lfvmajorana}.
This contribution to $(g-2)$ can be estimated as \cite{Lindner:2016bgg},
\begin{eqnarray}
    \label{eq:g-2Majorana}
    \Delta a_{\mu}=\frac{m_{\mu}^2}{(4\pi)^2}\sum_{i=1}^3(\lambda_{\psi_{\mu i}}^*\lambda_{\psi_{\mu i}})\int_{0}^1 dx \frac{(1-x)^2(x+\frac{m_{\psi_i^-}}{m_\mu})}{(1-x)(m_{\psi_i^-}^2 -x\, m_{\mu}^2)+x\,m_{\phi}^2}
\end{eqnarray}
The present value of the muon anomalous magnetic moment is reported to be $\Delta{a_{\mu}}=38(63)\times10^{-11}$ \cite{Muong-2:2025xyk,Aliberti:2025beg}. This indicates that, within the current uncertainties, there is no statistically significant tension between the experimental measurement and the SM prediction. Nevertheless, we will use the upper limit on $\Delta{a_{\mu}}=101\times10^{-11}$ to constrain the model parameter space in the later sections.

\subsection{Charged lepton flavor violation}
\label{subsec:lfvMajorana}

The observation of neutrino oscillations demonstrates that lepton flavor is not an exact symmetry of the Standard Model (SM). The charged lepton flavor–violating (cLFV) processes can occur in the SM at the one-loop level, although they are extremely suppressed by the tiny neutrino masses. In our framework, the presence of the singlet scalar ($\phi$) and the doublet fermions ($\Psi_i,\,i=1,2,3$) leads to additional cLFV contributions through the one-loop mediated diagrams shown in figure \ref{fig:lfvmajorana}. From there, one can find the branching ratio for the process $\mu\rightarrow e\gamma$ as given by \cite{Lindner:2016bgg},
\begin{eqnarray}
    \label{eq:LFVMajorana}
    {\rm Br}(\mu \rightarrow e\gamma)&\approx&  \frac{3(4\pi)^3 \alpha_{em}}{4 G_F^2}\times \Bigg\lvert\sum_{i=1}^3 \frac{\lambda_{\psi_{\mu i}}\lambda_{\psi_{e i}}^*}{(4\pi)^2} \int_0^1 dx \int_{0}^{1-x}dy \nonumber\\&&\frac{x(y+(1-x-y)\frac{m_e}{m_\mu})+(1-x)\frac{m_{\psi_i^-}}{m_\mu}}{-x\,y\,m_{\mu}^2 -x(1-x-y)m_{e}^2+x m_{\phi}^2 +(1-x)m_{\psi_i^-}^2} \Bigg\rvert^2 
\end{eqnarray}
The most recent constraint from the MEG-II collaboration sets an upper limit of ${\rm Br}(\mu\rightarrow e\gamma)<3.1\times10^{-13}$ at 90\% C.L. \cite{MEGII:2023ltw}. We impose this bound on our parameter space to ensure consistency with the desired phenomenology.
\begin{figure}[h]
    \centering \includegraphics[width=0.5\textwidth]{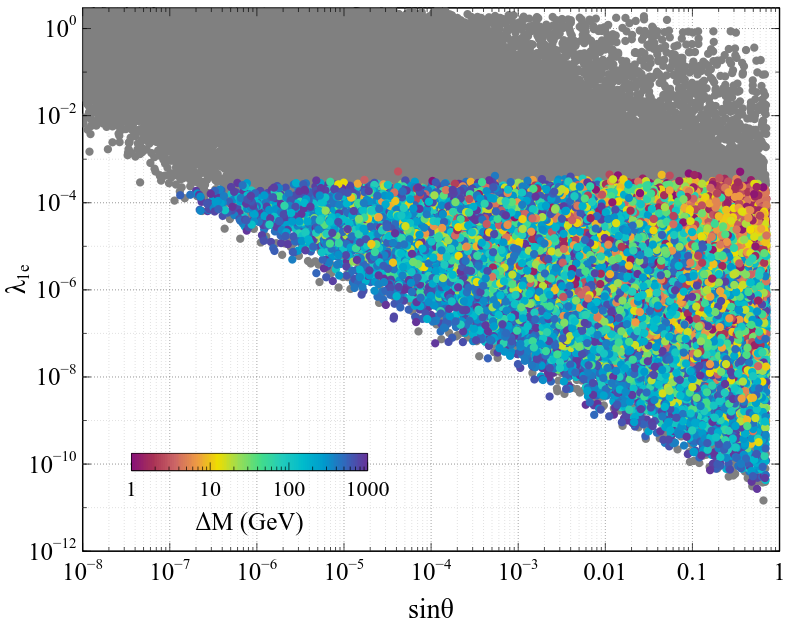}
    \caption{Parameter space consistent with neutrino mass, muon ($g-2$), and cLFV constraints is shown with colored points in the plane of $\lambda_{1e}$ vs $\sin\theta$. The color code depicts the value of the singlet-doublet mass splitting $\Delta{M}$. The gray points in the background satisfy only the neutrino mass.}
\label{fig:couplingvsangle}
\end{figure}
In figure~\ref{fig:couplingvsangle}, we show the parameter space consistent with neutrino mass, the muon anomalous magnetic moment $(g-2)\mu$, and cLFV constraints, by colored points in the $
\lambda_{1e}$–$\sin\theta$ plane, where the $\sin\theta$ represents the singlet-doublet mixing as given in the appendix~\ref{app:majo}. The color scale indicates the singlet–doublet mass splitting $\Delta M$. Gray points in the background satisfy the neutrino mass constraint only. We observe that the minimum allowed value of $\lambda_{1e}$ decreases with increasing $\sin\theta$, reflecting the seesaw structure underlying neutrino mass generation. As 
$\sin\theta$ increases, the loop factor increases, allowing smaller values of $\lambda_{1e}$ to reproduce the observed neutrino masses. However, the upper boundary of $\lambda_{1e}\lesssim \mathcal{O}(10^{-4})$ remains approximately constant across the entire range of $\sin\theta$, being set by the MEG-II bound on ${\rm Br}(\mu\rightarrow e\gamma)$. Consequently, for $\lambda_{1e}\sim\mathcal{O}(10^{-4})$, the minimum allowed value of $\sin\theta$ which gives correct neutrino mass is $\mathcal{O}(10^{-7})$. A similar constraint can be obtained for $\lambda_{1\mu}$ coupling as a function of $\sin\theta$.
\begin{table}[h!]
\centering
\begin{tabular}{|c|c|}
\hline
Parameter & Scan Range\\
\hline
$M_{\chi_3}$ (GeV) & $[1,2000]$ \\
\hline
$M_{\chi_1} - M_{\chi_3}$ (GeV) & $[1,1000]$\\
\hline
$\log(\sin\theta) $ & $-7,-0.15$\\
\hline
$M_{\phi} - M_{\chi_1}$ (GeV) & $[1,1000]$\\
\hline
$M_{\Psi_2} - M_{\phi}$ (GeV) & $[1,1000]$\\
\hline
$M_{\Psi_3} - M_{\Psi_2}$ & $[1,1000]$ \\
\hline
$M_{N_2} - M_{\Psi_3}$ (GeV) & $[1,1000]$\\
\hline
$M_{N_3}/M_{N_2}$ & $[1,1000]$ \\
\hline
$\log(\alpha_i,\beta_i)$ & $-10,5$ \\
\hline
\end{tabular}
\caption{Scan ranges of free parameters in the Majorana case.}
\label{tab:majorana_scan}
\end{table}
We vary the free parameters in the ranges as listed in table \ref{tab:majorana_scan}. We assume the following texture for the $N\Psi H$ coupling:
\begin{eqnarray}
    \label{eq:Yukawatexture}
    y=
    \begin{pmatrix}
        y_{11} & 0 & 0\\
        0 & y_{22} & y_{23}\\
        0 & y_{32} & y_{33}
    \end{pmatrix}.
\end{eqnarray}
The rationale for this texture is as follows. For simplicity, we choose to forbid the mixing of the heavier generations of the singlet–doublet fermions with the first generation\footnote{This texture can be realized with the help of of an additional $\mathcal{Z}^\prime_2$ symmetry, under which $\{\Psi_1,N_1\}$ are odd while $\{\Psi_{2,3},N_{2,3}\}$ are even.}. This choice prevents the decay of the heavier singlet generations into the DM. However, they can decay to charged SM leptons and $\phi$. In this setup, DM is produced through the first-generation Yukawa coupling $y_{11}$, while the second and third generation couplings are responsible for leptogenesis. In our analysis, we assume a hierarchy among the Yukawa couplings \textit{i.e.}, $y_{22}, y_{32} \ll y_{23}, y_{33}$ such that correct leptogenesis can be achieved in the sub-TeV regime while satisfying the constraints from cLFV and $(g-2)_{\mu}$ \footnote{However, if this condition is relaxed, leptogenesis can still be achieved but at a cost of relatively high scale.}.  We will discuss these aspects in detail in Sec.~\ref{sec:majoranalepto}.
\subsection{Dark matter phenomenology}\label{subsec:DM_A}
\subsubsection{Thermal relic of dark matter}\label{subsec:relicA}

\begin{figure}[h]
\centering
\includegraphics[width=0.475\textwidth]{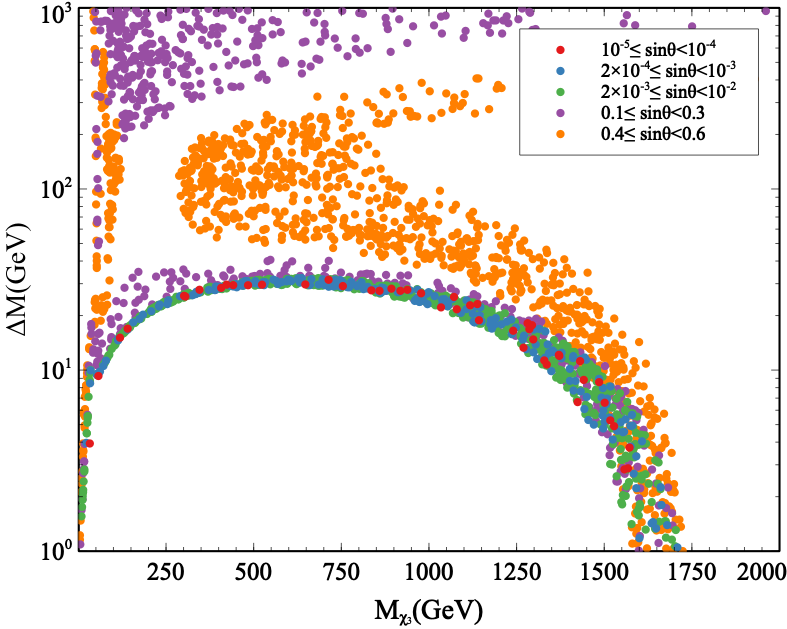}
\includegraphics[width=0.475\textwidth]{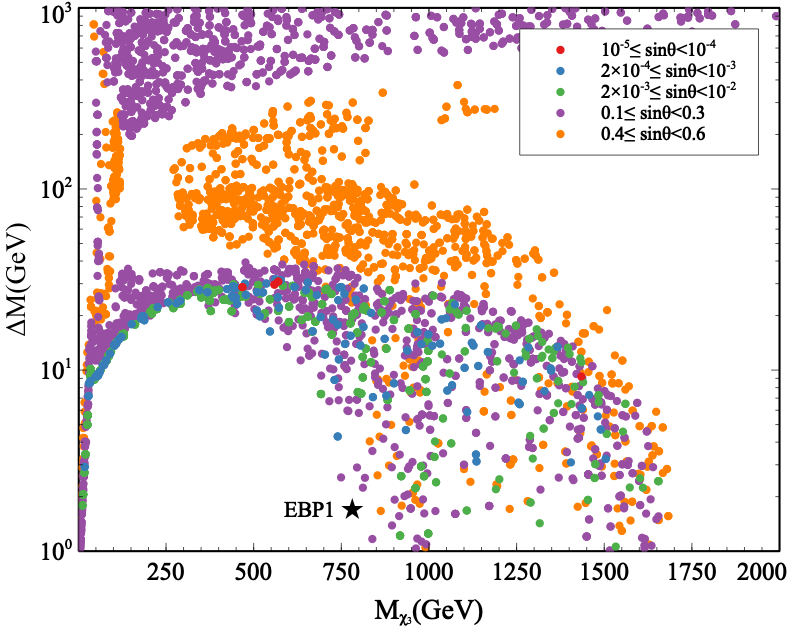}
\caption{[\textit{Left}]: Correct DM relic parameter space in the plane of $\Delta{M}$ vs $M_{\rm{DM}}$ considering all other particles to be very heavy compared to the first generation of singlet-doublet fermions. \textit{Right:} Plot for correct relic in the plane of $M_{\chi_3}$ vs $\Delta M=M_{\chi_1}-M_{\chi_3}$ with sine of the mixing angle $\rm{sin}\theta$ in the color code.}
\label{fig:correctrelicMajoranaDM}
\end{figure}

As we discussed in Sec.~\ref{subsec:modelA} $\chi_3$ is the lightest stable particle and serves as the Majorana singlet-doublet DM with mass $M_{\chi_3}\equiv M_{\rm{DM}}$, which is predominantly singlet-like. The details of the mixing are given in the Appendix \ref{app:majo}. In our analysis, we define two dark sectors: (a) sector 1, containing $\chi_3$, and (b) sector 2, comprising $\chi_1,~\chi_2$, $\psi_1^\pm$ and other additional particles ($N_{2,3},\,\Psi_{2,3}$ and $\phi$), while all SM particles are assigned to sector 0. We define the comoving number densities of sector 1 and sector 2 particles as $Y_1 \equiv n_{\chi_3}/s$ and $Y_2=(n_{\chi_1}+n_{\chi_2}+n_{\psi_1^\pm}+n_{N_{2,3}}+n_{\Psi_{2,3}}+n_\phi)/s$, respectively. The coupled Boltzmann equations governing their evolution are given by Eqs.~\ref{eq:Y1} and \ref{eq:Y2}, so that the total DM relic is $Y=Y_1+Y_2$. We'll see that $Y_2\ll Y_1$, therefore for all practical purposes $Y\approx Y_1$. We use \texttt{micrOMEGAs} \cite{Alguero:2022inz} to compute the relic density of DM. In our analysis, we explicitly verify the thermalization of the DM candidate and compute the relic abundance by taking into account not only annihilation and co-annihilation processes, but also conversion-driven processes as outlined in Ref.~\cite{Paul:2024prs,Paul:2025spm,DAgnolo:2017dbv,Garny:2017rxs}. The independent parameters relevant for the relic density computation are the DM mass $M_{\rm DM}$, the singlet–doublet mass splitting $\Delta{M}$, and the singlet–doublet mixing angle $\sin\theta$. We vary these parameters independently and compute the DM relic abundance. In figure \ref{fig:correctrelicMajoranaDM}, we present the region of parameter space consistent with the observed relic density in the $\Delta{M}$–$M_{\rm DM}$ plane. The color code represents the value of the mixing angle $\sin\theta$. We first considered a case where the second and third generations of the singlet-doublet, along with the singlet scalar, are much heavier compared to the first generation of singlet-doublet (\textit{left} panel of figure~\ref{fig:correctrelicMajoranaDM}). And in the second case, we considered a more general mass spectrum in which the other dark sector states can be close to the DM (\textit{right} panel of figure~\ref{fig:correctrelicMajoranaDM}).  
\begin{figure}[h]
\centering
\includegraphics[width=0.475\textwidth]{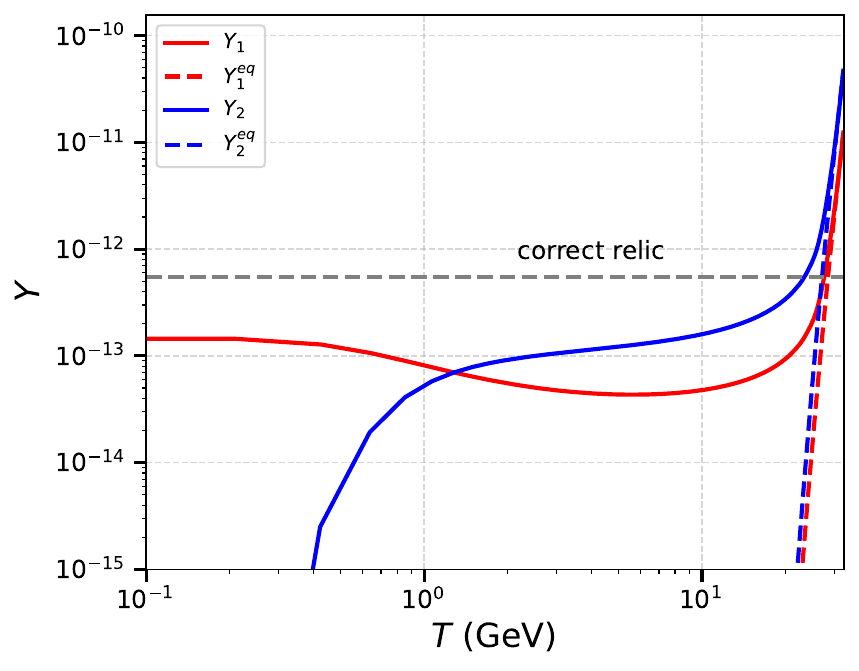}
\includegraphics[width=0.475\textwidth]{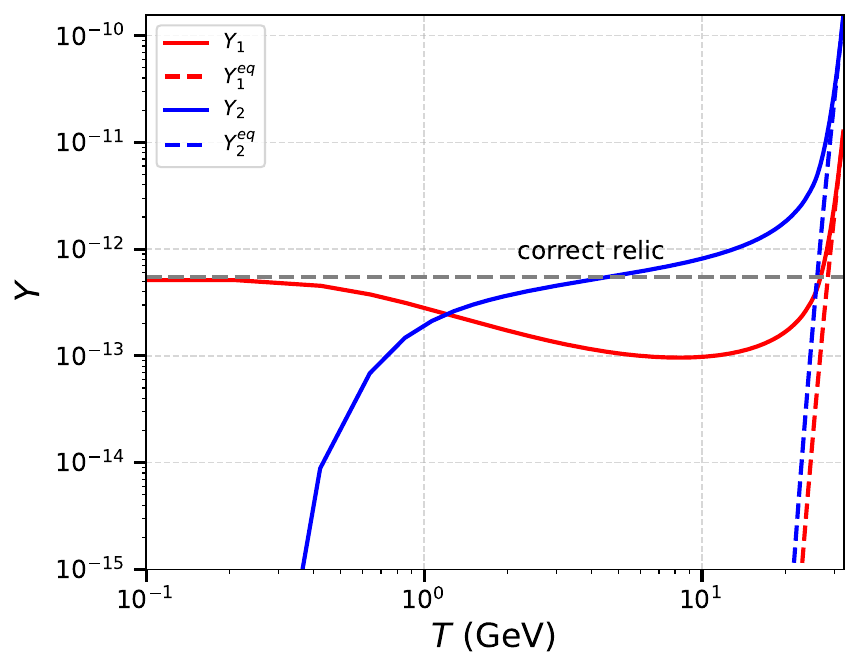}   \caption{[\textit{Left}]:  Cosmological evolution of abundances of sector 1 and sector 2 particles for the benchmark point in the decoupling limit. \textit{Right:} Cosmological evolution of abundances of sector 1 and sector 2 particles for the benchmark point when the masses of all particles are close to each other.}
\label{fig:MajoranaDM_evo}
\end{figure}

As the DM mass increases, the annihilation cross-section decreases, thereby enhancing the relic abundance. Coannihilation effects play a crucial role in reducing the relic density to the observed value, particularly when the mass splitting between the DM and the next-to-lightest stable particle (NLSP) is small. This behavior is clearly illustrated in figure~\ref{fig:correctrelicMajoranaDM} (left). For large mass splittings $\Delta M$, coannihilation becomes negligible, and the relic density is primarily governed by Higgs-mediated annihilation processes. In this regime, the mixing angle $\sin\theta$ becomes important, as the relevant Yukawa coupling scales as $\propto \Delta M \sin 2\theta$. For a fixed value of $\sin\theta$, increasing $\Delta M$ enhances the coupling, thereby increasing the annihilation cross-section and reducing the relic abundance. Consequently, achieving the correct relic density requires a larger DM mass, which appropriately suppresses the cross-section. This mimics the pure singlet-doublet Majorana DM scenario \cite{Bhattacharya:2018fus,Dutta:2020xwn,Paul:2025spm}. 

In the \textit{right} panel of figure \ref{fig:correctrelicMajoranaDM}, we show the correct relic satisfying points in the same plane where the other dark sector states ($N_2,\Psi_2,N_3,\Psi_3,\phi$) can be close to DM also. It is interesting to note that the correct DM relic parameter space gets significantly modified compared to the pure singlet-doublet Majorana DM \cite{Paul:2025spm}, as shown in the \textit{left} panel. In the decoupling limit (\textit{left} panel of figure \ref{fig:correctrelicMajoranaDM}), the DM remains under-abundant for $M_{\chi}$ near 800 GeV to 1000 GeV with $\Delta{M}\lesssim20$ GeV and $\sin\theta\gtrsim10^{-5}$ due to large co-annihilation and conversion driven processes. However, in the presence of second and third generations of singlet-doublet fermions and singlet scalar, we get correct DM relic points in this ``previously" under-abundant region. This is mainly due to the presence of additional dark sector states, which contribute to the final relic of DM.
\begin{table*}[t]
    \centering
    \setlength{\tabcolsep}{6pt}
    \renewcommand{\arraystretch}{1}
    \begin{tabular}{|l|c|c|c|}
    \hline
        Input parameters & EBP1 & EBP2\\
        \hline
        $M_{\chi_3}\,(\rm{GeV})$ & $781.332$ & $781.332$\\
        \hline
        $\Delta{M}\,(\rm{GeV})$ & $1.697$ & $1.697$\\
        \hline
        $\sin\theta$ & $0.03397$ & $0.03397$\\
        \hline
        $M_{\Psi_2}\,(\rm{GeV})$ & $785.352$ & $7085.352$\\
        \hline
        $M_{\Psi_3}\,(\rm{GeV})$ & $789.942$ & $7089.942$\\
        \hline
        $M_{\phi}\,(\rm{GeV})$ & $785.231$ & $7085.231$\\
        \hline
        $M_{N_2}\,(\rm{GeV})$ & $791.764$ & $7091.764$\\
        \hline
        $M_{N_3}\,(\rm{GeV})$ & $800.701$ & $8000.701$\\
        \hline
        $\theta_1$ & $-2.288\times10^{-6} + 3.379\times10^{-10}\,\iota$ & $-2.288\times10^{-6} + 3.379\times10^{-10}\,\iota$\\
        \hline
        $\theta_2$ & $1.739\times10^{-9} + 9.199\times10^{-7}\,\iota$ & $1.739\times10^{-9} + 9.199\times10^{-7}\,\iota$\\
        \hline
        $\theta_3$ & $-4.215\times10^{-9} - 1.132\times10^{-5}\,\iota$ & $-4.215\times10^{-9} - 1.132\times10^{-5}\,\iota$\\
        \hline
    \end{tabular}
    \caption{Input parameters for the two benchmark points (EBP1, EBP2) used in figure \ref{fig:MajoranaDM_evo}.}
\label{tab:DarkmajoranBP}
\end{table*}
To understand the effect of additional dark sector particles on the DM relic, we choose a benchmark point (BP) EBP1, indicated by the black star, from the \textit{right} panel of figure~\ref{fig:correctrelicMajoranaDM}.
The details of the EBP1 are given in Table~\ref{tab:DarkmajoranBP}. The Yukawa coupling matrix $y$ for EBP1 is given as:
\begin{eqnarray}
    y=
    \begin{pmatrix}
        3.312\times10^{-4} & 0 & 0\\
        0 & 7.686\times10^{-1} & 3.750\times10^{-2}\\
        0 & 1.164\times10^{-1} & 2.052\times10^{-1}
    \end{pmatrix},
\end{eqnarray}
and the $\bar{L}\Psi\phi$ coupling is given as:
\begin{eqnarray}
    \lambda=
    \begin{pmatrix}
-4.47\times10^{-9} - 3.21\times10^{-9} \iota &
-1.91\times10^{-7} + 1.41\times10^{-6} \iota &
-1.59\times10^{-6} - 1.53\times10^{-5} \iota \\

5.99\times10^{-9} + 7.69\times10^{-9} \iota &
-1.27\times10^{-7} - 2.77\times10^{-6} \iota &
-1.06\times10^{-6} + 1.73\times10^{-5} \iota \\

-7.84\times10^{-9} + 7.65\times10^{-9} \iota &
-8.33\times10^{-11} - 2.88\times10^{-6} \iota &
1.07\times10^{-10} - 2.40\times10^{-5} \iota
\end{pmatrix}\nonumber\\
\end{eqnarray}
In the \textit{right} panel of figure~\ref{fig:MajoranaDM_evo}, we show the evolution of the particle abundances of sector 1 and sector 2 as a function of temperature. The abundance of sector 1 ($Y_1$) is shown by the red solid line, while the red dashed line represents its equilibrium abundance. The blue solid and dashed lines correspond to the abundance of sector 2 particles ($Y_2$) and their equilibrium values, respectively. Since the masses of the 2nd and 3rd generations of the singlet-doublet fermions and $\phi$ are close to that of the first-generation singlet-doublet fermion, the abundances of the other dark sector particles are efficiently converted into the first-generation doublet during the epoch when both the DM and sector 2 particles decouple. As a result, the sector 2 abundance is enhanced, and at later times, this excess abundance is transferred to sector 1. Consequently, the DM eventually freezes out with the correct relic abundance.

We now make the 2nd and 3rd generations of the singlet-doublet fermions and the singlet scalar much heavier than the first generation, such that the scenario effectively mimics a pure singlet-doublet DM framework. The details of the EBP2 are also given in Table~\ref{tab:DarkmajoranBP}. The Yukawa coupling matrix $y$ for EBP2 is given as:
\begin{eqnarray}
    y=
    \begin{pmatrix}
        3.312\times10^{-4} & 0 & 0\\
        0 & 7.686\times10^{-1} & 3.750\times10^{-2}\\
        0 & 1.164\times10^{-1} & 2.052\times10^{-1}
    \end{pmatrix},
\end{eqnarray}
and the $\bar{L}\Psi\phi$ coupling is given as:
\begin{eqnarray}
    \lambda=
    \begin{pmatrix}
-1.34\times10^{-8} - 9.64\times10^{-9} \iota &
-5.85\times10^{-7} + 4.32\times10^{-6} \iota &
-4.92\times10^{-6} - 4.71\times10^{-5} \iota \\

1.80\times10^{-8} + 2.31\times10^{-8} \iota &
-3.89\times10^{-7} - 8.49\times10^{-6} \iota &
-3.28\times10^{-6} + 5.35\times10^{-5} \iota \\

-2.35\times10^{-8} + 2.29\times10^{-8} \iota &
-2.55\times10^{-10} - 8.81\times10^{-6} \iota &
3.29\times10^{-10} - 7.41\times10^{-5} \iota
\end{pmatrix}\nonumber\\
\end{eqnarray}
The evolution of the abundances of $Y_1$ and $Y_2$ for this BP is shown in the \textit{left} panel of figure~\ref{fig:MajoranaDM_evo}. Since the 2nd and 3rd generations are much heavier than the first generation, the abundances of $\Psi_2, N_2, \Psi_3, N_3,$ and $\phi$ are transferred to the first generation doublets, which get thermalized quickly. As a result, no enhancement in the abundance of $Y_2$ is observed when it decouples from the thermal bath. In this case, the sector 2 abundance is much smaller than the sector 2 abundance of EBP1, \textit{i.e.} $Y^{\rm EBP2}_{2}\ll Y^{\rm EBP1}_{2}$. Consequently, the DM remains underabundant. This highlights the impact of other dark sector particles on the singlet-doublet DM relic.
\subsubsection{Direct detection prospects}\label{subsec:DD_A}

\begin{figure}[h]
    \centering
    \includegraphics[width=0.475\textwidth]{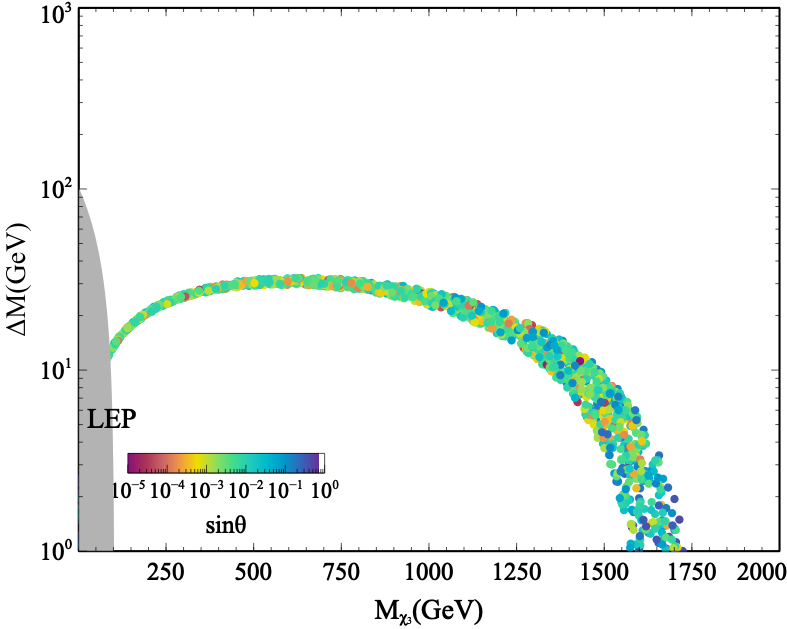}
    \includegraphics[width=0.475\textwidth]{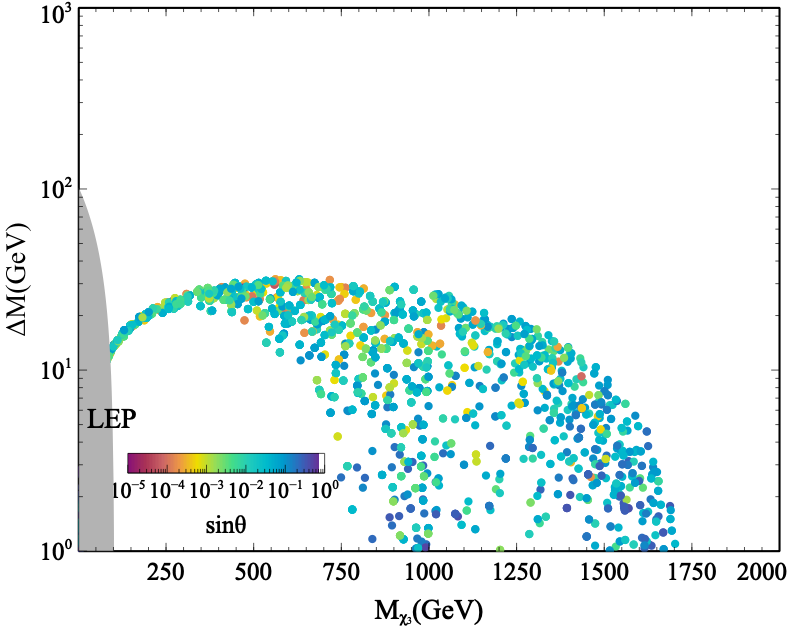}
    \caption{[\textit{Left}]: points satisfying correct relic and direct detection from the LZ experiment are shown in the $\Delta M -M_{\rm DM}$ plane for the decoupling limit. The color code represents $\rm{sin}\theta$. [\textit{Right}]: same as the \textit{left} but for more general case. Note that all these points satisfy the constraints from neutrino mass, $(g-2)$, and cLFV.}
\label{fig:crctrelafterDD}
\end{figure}

In this scenario, direct detection is possible via Higgs exchange. We calculate the spin-independent direct detection cross-sections for the points shown in figure \ref{fig:correctrelicMajoranaDM} and impose the constraint from  LZ \cite{LZ:2024zvo}. The allowed points from direct detection are shown in the plane of $M_{\rm DM}$ and $\Delta{M}$ in figure \ref{fig:crctrelafterDD}. The \textit{left} plot corresponds to the decoupling limit (the second and third generation of singlet-doublet fermions $N_{2,3}, \,\Psi_{2,3}$ and the singlet scalar $\phi$ to have mass much larger than the mass of first generation of singlet-doublet fermions), and the \textit{right} one corresponds to the general case, where the masses and the mass splittings are varied in accordance to Table~\ref{tab:majorana_scan}. The allowed $\sin\theta$ in both cases are in similar range of $\mathcal{O}(10^{-5})$ to $0.484$.
\subsection{Leptogenesis from $N_{2,3}$ decay}
\label{sec:majoranalepto}

\begin{figure}[h]
    \centering
    \includegraphics[scale=0.8]{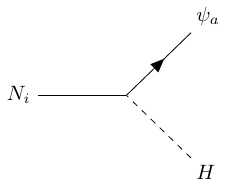}
    \includegraphics[scale=0.8]{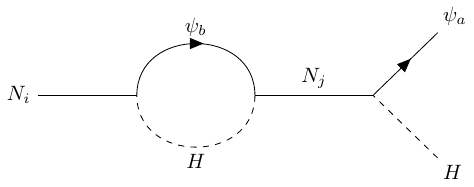}
    \includegraphics[scale=0.8]{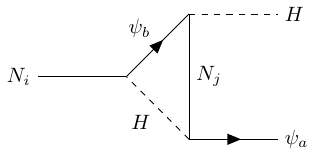}
    \caption{Tree level and one loop diagrams contributing to non-zero \textit{CP} asymmetry.}
    \label{fig:majoranaCP}
\end{figure}

In the early universe, the \textit{CP}-violating out-of-equilibrium decays of heavier generations of singlet fermions ($N_2$, $N_3$) can generate asymmetries in the $\Psi_2,\Psi_3$ through the decay channel $N\rightarrow\Psi H$. This asymmetry is subsequently transferred to the SM leptons through the decay processes $\Psi_{2,3} \rightarrow L \phi$. The first-generation singlet–doublet fermions, being the lightest, do not participate in leptogenesis. The resulting lepton asymmetry is partially converted into the baryon asymmetry via electroweak sphaleron processes. The decay width of the singlet fermions $N_{i},$ is given by
\begin{eqnarray}
\label{decayrateNi}
    \Gamma_{D_{N_i}}(z)&=&\frac{(y^\dagger y)_{ii}}{8\pi} {(M_{N_i} + M_{\Psi_i})^2}\left(1-\frac{M_{\Psi}^2}{M_{N_i}^2}\right)\frac{K_1\left(\frac{M_{N_i}}{M_{N_2}} z\right)}{K_2\left(\frac{M_{N_i}}{M_{N_2}} z\right)}.
\end{eqnarray}
The CP asymmetry parameter arising from the interference of the tree-level and one-loop decay of $N_i \longrightarrow\Psi_i \, \phi$, as shown in figure \ref{fig:majoranaCP}, is given by 
\begin{eqnarray}
\label{CPasymmetry}
    \epsilon_{i} = \frac{1}{8\pi} \frac{1}{(y^\dagger y)_{ii}}\sum_{j\neq i} {\rm Im}[(y^\dagger y)^2_{ji}] f\left(\frac{M_j^2}{M_i^2}\right)
\end{eqnarray}
where,
\begin{eqnarray}
    \label{loopfunc}
    f(x) = \sqrt{x}\left(\frac{1}{1-x}+1-(1+x)\ln\left(\frac{1+x}{x}\right)\right)
\end{eqnarray}

The decay width of the doublet fermion $\Psi_i$ to $L\phi$ is given as
\begin{eqnarray}
\label{decaywidthpsii}
    \Gamma_{D_{\Psi_i}} = \frac{(\lambda^\dagger\lambda)_{ii}M_{\Psi_i}}{16\pi}\left(1-\frac{M_\phi^2}{M_{\Psi_i}^2}\right)^2\frac{K_1\left(\frac{M_{\Psi_i}}{M_{N_2}}z\right)}{K_2\left(\frac{M_{\Psi_i}}{M_{N_2}}z\right)}
\end{eqnarray}

To track the abundances of $N_{2,3}$ and the asymmetries in the  doublet ($Y_{\Delta\Psi}$) and lepton sector ($Y_{\Delta{L}}$), we solve the following set of Boltzmann equations
\begin{eqnarray}
    \frac{dY_{N_i}}{dz}&=&-\frac{z}{\mathcal{H}(M_{N_2})}\Bigg[\Gamma_{D_{N_i}}\left(Y_{N_i}(z) - Y_{N_i}^{eq}(z)\right) + \sum_{j=2,3}n_\gamma(z)\bigg[ \left(Y_{N_i}(z)Y_{N_j}(z) - Y_{N_i}^{eq}(z)Y_{N_j}^{eq}(z)\right)\times\nonumber\\&& \langle\sigma_{N_i N_j\longrightarrow HH}\rangle \bigg] +\sum_{j=1}^3 n_\gamma(z)\bigg[\left(Y_{N_i}(z)-Y_{N_i}^{eq}(z)\right)Y_{\Psi_j}^{eq}(z) \bigg(\langle \sigma_{N_i \Psi_j\rightarrow ZZ}\rangle+ \langle \sigma_{N_i \Psi_j\rightarrow WW}\rangle\times\nonumber\\&&+\langle\sigma_{N_i \Psi_j \longrightarrow t T}\rangle\bigg) + \left(Y_{N_i}(z)- Y_{N_i}^{eq}(z)\right)  \, Y_{Z}^{eq}(z)\left(\langle\sigma_{N_i Z \longrightarrow \Psi_j Z}\rangle+\langle\sigma_{N_i W \longrightarrow \Psi_j W}\rangle\right)\nonumber\\&& + \left(Y_{N_i}(z)- Y_{N_i}^{eq}(z)\right) Y_{\phi}^{eq}(z)\langle\sigma_{N_i \phi \longrightarrow \nu_j H}\rangle + \left(Y_{N_i}(z)- Y_{N_i}^{eq}(z)\right) Y_l^{eq} \langle\sigma_{N_i \nu_j \longrightarrow \phi H}\rangle\bigg]\Bigg]
\end{eqnarray}
\begin{eqnarray}
    \frac{dY_{\Delta\Psi}}{dz} &=& -\frac{z}{\mathcal{H}(M_2)} \left(\sum_{i=2,3}\epsilon_i \Gamma_{D_{N_i}} (Y_{N_i} - Y_{N_i}^{eq}) + \Gamma_{D_{\Psi_i}}(Y_{\Delta \Psi} - \frac{Y_{\Psi_i}^{eq}}{Y_l^{eq}}Y_{\Delta L})  + \frac{1}{2} \Gamma_{D_{\Psi_i}}\frac{Y_{N_i}^{eq}}{Y_{\Psi_i}^{eq}}Y_{\Delta\Psi}\right) \nonumber\\ &&+ s\left( \Gamma_{\Psi\Psi\longrightarrow HH} + \Gamma_{\Psi H\longrightarrow \Psi H} + \Gamma_{ N\Psi\longrightarrow t Q}\right)Y_{\Delta\Psi}
\end{eqnarray}
\begin{eqnarray}
    \frac{dY_{\Delta L}}{dz} = \frac{z}{\mathcal{H}(M_{N_2})} \left( \Gamma_{D_{\Psi_i}} Y_{\Delta\Psi} - \Gamma_{D_{\Psi_i}} \frac{Y_{\Psi_i}}{Y_l^{eq}}Y_{\Delta L} - s(\Gamma_{N L\longrightarrow H \phi} + \Gamma_{H L\longrightarrow N \phi})Y_{\Delta L} \right),
\end{eqnarray}
where $z = \frac{M_{N_2}}{T}$, and $Y_x$ denotes the comoving number density of the particle species $x$, defined as $Y_x = n_x / n_\gamma$, with $n_x$ being the number density of $x$ and $n_\gamma$ the photon number density. Here, $\mathcal{H}$ represents the Hubble parameter. Above the electroweak phase transition, the net lepton asymmetry produced can be converted to the observed baryon asymmetry via nonperturbative sphaleron processes. The resulting baryon asymmetry can be given as,
\begin{eqnarray}
\eta_B=\frac{C}{f}Y_{\Delta L},
\end{eqnarray}
where $f$ = 33.8\footnote{$g_{*s}=106.75+\frac{7}{8}(2+2+(4+4)\times3)+1=132.25$, is the relativistic d.o.f at the onset of leptogenesis and $g_{*0}$= 3.91 is the relativistic d.o.f today. The dilution factor f is then,$f=g_{*s}/g_{*0}=33.8$.} is the dilution factor calculated assuming standard photon production from the onset of leptogenesis till recombination. $C$ is the lepton to baryon asymmetry conversion factor, and in our case, the value of it is $C=-0.527$\footnote{See appendix \ref{app:ltob_maj} for more details.}.

\begin{table*}[t]
    \centering
    \setlength{\tabcolsep}{6pt}
    \renewcommand{\arraystretch}{1}
    \begin{tabular}{|l|c|c|c|}
    \hline
        Input parameters & MBP1 & MBP2\\
        \hline
        $M_{\chi_3}\,(\rm{GeV})$ & $157.81$ & $157.81$\\
        \hline
        $M_{\chi_2}\,(\rm{GeV})$ & $175.90$ & $175.90$\\
        \hline
        $M_{\chi_1}\,(\rm{GeV})$ & $175.90$ & $175.90$\\
        \hline
        $\sin\theta$ & $0.01$ & $0.01$\\
        \hline
        $M_{\Psi_2}\,(\rm{GeV})$ & $750.84$ & $465.46$\\
        \hline
        $M_{\Psi_3}\,(\rm{GeV})$ & $833.61$ & $610.65$\\
        \hline
        $M_{\phi}\,(\rm{GeV})$ & $586.36$ & $460.29$\\
        \hline
        $M_{N_2}\,(\rm{GeV})$ & $900$ & $650.41$\\
        \hline
        $M_{N_3}\,(\rm{GeV})$ & $9.00\times10^{7}$ & $6.50\times10^{7}$\\
        \hline
        $\theta_1$ & $2.364\times10^{-5}-2.065\times10^{-3}\iota$ & $-4.14 \times 10^{-4}+4.38\times 10^{-2}\iota$\\
        \hline
        $\theta_2$ & $-5.268\times 10^{-2}+4.442\times 10^{-5}\iota$ & $-1.58\times 10^{-5}+7.92\times10^{-5}\iota$\\
        \hline
        $\theta_3$ & $-1.315 + 2.07\times 10^{-5}\iota$ & $-3.7\times 10^{-1}+5.6\times 10^{-2}\iota$\\
        \hline
    \end{tabular}
    \caption{Input parameters for the two benchmark points used in Majorana leptogenesis.}
\label{tab:Majoranatable}
\end{table*}
We now solve the Boltzmann equations for two benchmark points, MBP1 and MBP2, as given in table \ref{tab:Majoranatable}, which are selected from the above analysis that satisfy the constraints from neutrino masses, muon $g-2$, cLFV, the observed DM relic density, and direct detection. The MBP1 leads to the following Yukawa coupling matrix\footnote{It is worth noting that in the presence of non-zero $y_{12},y_{13},y_{21}$ and $y_{31}$ couplings, the 2nd and 3rd generation singlets can decay to the first generation doublets via $N_{2,3}\rightarrow \Psi_1 H$. This generates a non-zero asymmetry in $\Psi_1$, which subsequently gets converted to the DM $N_1$. Thus giving rise to an asymmetric DM relic. However, as the DM is Majorana in our setup, the asymmetry vanishes.}

\begin{eqnarray}
\label{eq:bp01yukawa}
    y=\begin{pmatrix}
    1.04\times10^{-3} & 0 & 0\\
    0& -5.36\times10^{-7}+6.59\times10^{-7}\iota & -5.08 \times 10^{-2}-1.91 \iota\\
    0& 1.828\times 10^{-7}-1.854\times 10^{-8}\iota &-7.41\times 10^{-2}+1.098\iota
    \end{pmatrix}.
\end{eqnarray}

The \textit{CP} asymmetry parameters are found to be $\epsilon_2=-2.512\times10^{-6},\epsilon_3=5.473\times10^{-18}$. We compute the decay parameters\footnote{Decay parameter is defined as $K=\frac{\Gamma(z=\infty)}{H(z=1)}$. A value of $K\ll1$ indicates weak washout, and $K\gg1$ indicates strong washout of asymmetry.} for MBP1 and found to be $K_2=24.31,K_3=1.528\times10^9$. It is important to note that the \textit{CP} asymmetry due to $N_3$ is extremely small, while its corresponding decay parameter is very large. Consequently, the contribution of $N_3$ to the production of the $\Psi_3$ asymmetry is negligible. The primary effect of $N_3$ would be to wash out the asymmetry generated by $N_2$ due to its large decay parameter $K_3$. However, since $M_{N_3} = 9.00 \times 10^{7}$ GeV, $N_3$ decays well before $z \sim 10^{-5}$ and therefore does not significantly influence the evolution of $N_2$. For completeness, we nevertheless include $N_3$ in the Boltzmann equations, although it has no impact on the final asymmetry generated by $N_2$.
\begin{figure}[h]
    \centering
    \includegraphics[width=0.475\textwidth]{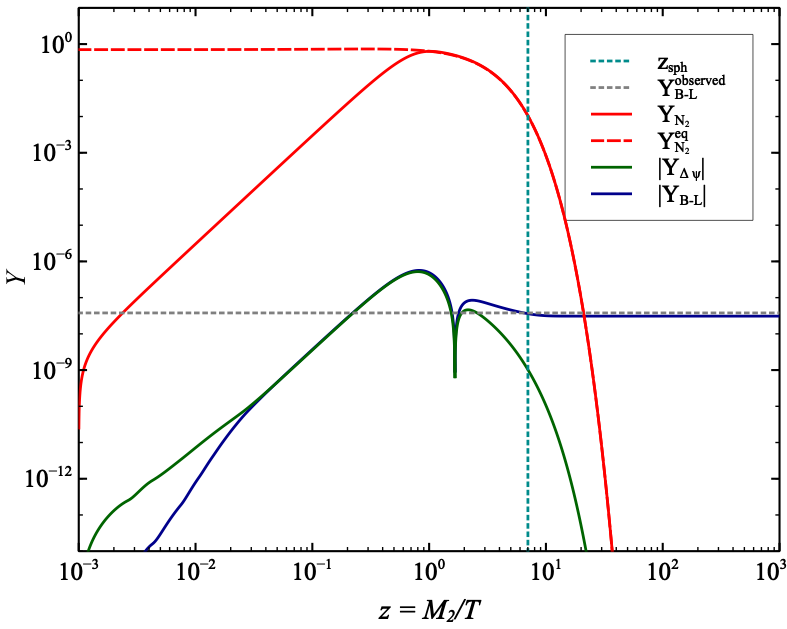}
    \includegraphics[width=0.475\textwidth]{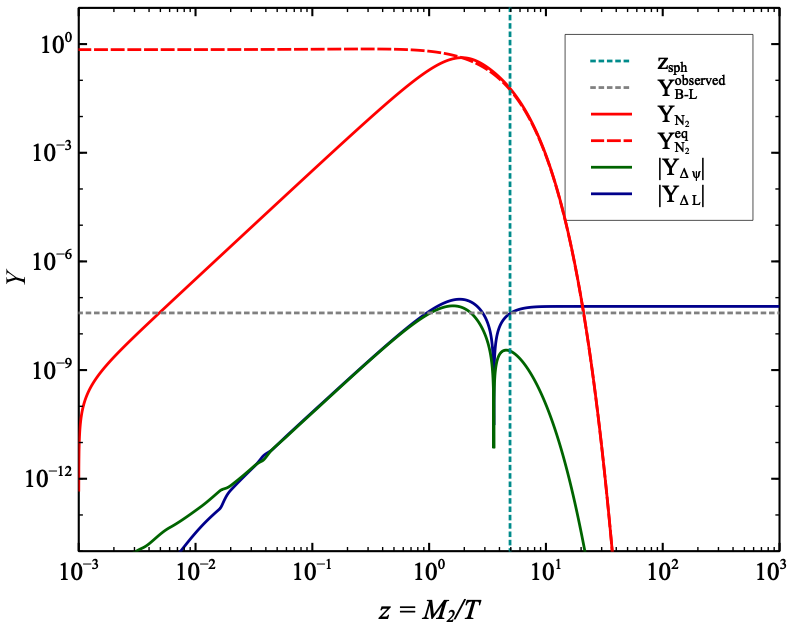}
    \caption{[\textit{Left}]: cosmological evolution of $N_2$ abundance, $\Psi$ asymmetry, and lepton asymmetry are shown with red, green, and dark-blue solid colored lines, respectively, for MBP1. The equilibrium abundance of $N_2$ is shown with a red dashed line. The dark-cyan vertical dotted line represents the sphaleron transition temperature, $T_{\rm sph}\simeq131.7$ GeV \cite{DOnofrio:2014rug}. The gray dotted horizontal line corresponds to the correct baryon asymmetry value. [\textit{Right}]: the same as in the \textit{left} panel, but for MBP2.}
\label{fig:Majoranabenchmarkpt}
\end{figure}
In the \textit{left} panel of figure \ref{fig:Majoranabenchmarkpt}, we show the evolution of the asymmetries as a function of $z=M_{N_2}/T$. The decay of $N_2$ first generates an asymmetry in the $\Psi$ sector, as indicated by the green solid line. This asymmetry is subsequently transferred to the lepton sector, shown by the dark blue solid line. The generated lepton asymmetry must be converted into a baryon asymmetry through electroweak sphaleron processes before the sphalerons freeze out. The vertical dark cyan dotted line denotes the sphaleron transition temperature, $T_{\rm sph}\simeq131.7$ GeV\footnote{The sphaleron decoupling temperature also does not change relative to the standard scenario. The Yukawa couplings $\Psi NH$ are small and do not give corrections to the thermal mass of Higgs. Therefore, the sphaleron rate and freeze-out temperature remain unchanged, and it is safe to use $T_{\rm sph}\simeq 131.7$ GeV for our phenomenological analysis.} \cite{DOnofrio:2014rug}. This benchmark point corresponds to the strong washout regime, and as a result, we observe a suppression in the final asymmetry. The resulting baryon asymmetry at $z_{\rm sph}=M_{N_2}/T_{\rm sph}$ is $\eta_B \sim 6.0 \times 10^{-10}$. 

In the \textit{right} panel of figure \ref{fig:Majoranabenchmarkpt}, we show the evolution of the asymmetries as a function of $z=M_{N_2}/T$ for the MBP2. This BP leads to the following Yukawa coupling matrix

\begin{eqnarray}
\label{eq:bp02yukawa}
    y=\begin{pmatrix}
    1.04\times10^{-3} & 0 & 0\\
    0& 1.584\times10^{-7}+1.185\times10^{-7}\iota & 9.064 \times 10^{-1}-1.110\iota\\
    0& 1.739\times 10^{-8}+4.69\times 10^{-8}\iota &-1.598\times 10^{-1}+1.054\iota
    \end{pmatrix}.
\end{eqnarray}

For MBP2, the \textit{CP} asymmetry parameters are $\epsilon_2 = -4.32 \times 10^{-7}$ and $\epsilon_3 = 7.92 \times 10^{-20}$, with corresponding decay parameters $K_2 = 2.60$ and $K_3 = 1.38 \times 10^9$. Similar to MBP1, the contribution of $N_3$ is negligible due to its extremely small \textit{CP} asymmetry and very large decay parameter. Although $N_3$ could, in principle, induce washout, its large mass ($M_{N_3} = 6.50 \times 10^{7}$ GeV) ensures that it decays well before $z \sim 10^{-5}$ and does not affect the evolution driven by $N_2$. This benchmark point corresponds to a relatively small washout regime, and therefore, no significant suppression of the final asymmetry is observed. The resulting baryon asymmetry at $z_{\rm sph}$ is $\eta_{B}\sim 6.0\times10^{-10}$. For these two benchmark points, the DM relic abundance is found to be $\Omega_{\rm DM} h^2 = 0.1159$, while the corresponding direct detection cross section is $\sigma = 5.103 \times 10^{-52}~\mathrm{cm}^2$. This cross-section lies well below the current sensitivity of the LZ experiment and falls below the neutrino floor. Thus, this BP is beyond the reach of current direct detection experiments, and the most promising probes of this scenario remain collider signatures.
\section{Model B: singlet-doublet Dirac dark matter and leptogenesis}\label{sec:model_B}

\subsection{The model}\label{subsec:model_B}
To realize the Dirac neutrino mass at one loop level, we extend the Standard Model by introducing a vector-like singlet fermion $\chi$, a vector-like fermion doublet $\Psi =(\psi^0~~\psi^-)^T$, three generations of complex scalars $\phi_i$, and three generations of Dirac right-handed neutrinos $\nu_{R_i}$ \cite{Borah:2023dhk}. A discrete $\mathcal{Z}_4$ symmetry is imposed under which the fermions $\chi$ and $\Psi$ carry charge $-1$, while the scalar fields $\phi_i$ carry charge $\mathbf{i}$. The Standard Model lepton doublets $L$ and the right-handed charged leptons $\ell_R$ transform with charge $-\mathbf{i}$, whereas the Dirac right-handed neutrinos $\nu_R$ have charge $\mathbf{i}$. The relevant Lagrangian terms and the interactions among the particles are given as
\begin{table}[h!]
		\small
		\begin{center}
			\begin{tabular}{||@{\hspace{0cm}}c@{\hspace{0cm}}|@{\hspace{0cm}}c@{\hspace{0cm}}|@{\hspace{0cm}}c@{\hspace{0cm}}|@{\hspace{0cm}}c@{\hspace{0cm}}||}
				\hline
				\hline
				\begin{tabular}{c}
                {\bf ~~~~Symmetry~~~~}\\
					{\bf ~~~~Group~~~~}\\ 
					\hline
					
					$SU(2)_{L}$\\ 
					\hline
					$U(1)_{Y}$\\ 
					\hline
					$Z_4$\\ 
				\end{tabular}
				&
				&
				\begin{tabular}{c|c|c|c|c}
					\multicolumn{5}{c}{\bf Fermion Fields}\\
					\hline
					~~~$L$~~~&~~~$\ell_R$~~~& ~~~$\Psi =(\psi^0~~\psi^-)^T $~~~ & ~~~$\chi$~~~&~~~$\nu_R$~~~ \\
					\hline
					$2$&$1$&$2$&$1$&$1$\\
					\hline
					$-1$&$-2$&$-1$&$0$&$0$\\
					\hline
					$-\mathbf{i}$&$-\mathbf{i}$&$-1$&$-1$&$\mathbf{i}$\\
				\end{tabular}
				&
				\begin{tabular}{c|c|c}
					\multicolumn{2}{c}{\bf Scalar Field}\\
					\hline
					~~~$H$~~~& ~~~$\phi_i$~~~\\
					\hline
					$2$&$1$\\
					\hline
					$1$&$0$\\
                    \hline
					$+1$&$\mathbf{i}$\\
				\end{tabular}\\
				\hline
				\hline
			\end{tabular}
			\caption{Particles and their charge assignments under the symmetry group $SU(2)_L\otimes U(1)_Y\otimes \mathcal{Z}_4$.}
			\label{tab1}
		\end{center}    
	\end{table}
\begin{eqnarray}
    \mathcal{L}\supset && \iota\bar{\Psi}\gamma^{\mu}D_{\mu}\Psi +\iota\bar{\chi}\gamma^{\mu}D_{\mu}\chi + (D_{\mu}\phi_i)^*(D^{\mu}\phi_i)  - M_{\Psi}\bar{\Psi}\Psi - M_{\chi} \bar{\chi}\chi - \lambda_{\Psi_{i \alpha}}\bar{L}_{\alpha}\phi_i \Psi  -  y\bar{\Psi}\tilde{H}\chi\nonumber\\&&- \lambda_{\chi_{i \beta}} \bar{\chi} \phi_i \nu_{R_{\beta}} + {\rm H.c.}-V(\phi_i,H),
\end{eqnarray}
where the most general scalar potential is given as
\begin{eqnarray}
    V(\phi,H)=&&-\mu_{H}^2(H^\dagger H)+\lambda_H (H^\dagger H)^2 +\lambda_{\phi H}(\phi_i^\dagger\phi_i)(H^\dagger H) + M_{\phi_{ii}}^2(\phi_i^\dagger\phi_i) +\lambda_{\phi_{ii}}(\phi_i^\dagger\phi_i)^2\nonumber\\&&+\lambda_{\phi_{ij}}(\phi_i^\dagger\phi_i)(\phi_j^\dagger\phi_j)
\end{eqnarray}
\
In order to generate the Dirac mass for neutrinos, the $Z_4$ symmetry must be broken. We achieve this by introducing a soft symmetry-breaking term in the scalar potential, $\frac{1}{2}\mu_{\phi_i}^2 \left(\phi_i^2 + (\phi_i^\dagger)^2 \right),$ which explicitly breaks the $Z_4$ symmetry\footnote{The $\mathcal{Z}_4$ symmetry forbids the $LH\nu_R$ term at tree level and neutrinos remain massless. A softly broken $\mathcal{Z}_4$ symmetry reintroduces this term at one loop, and neutrinos become massive without fine-tuning of the Yukawa couplings. One important consequence of generating neutrino mass at the loop level is that the dark sector particles involved in the loop lie at a low scale, ranging from the TeV to hundreds of TeV. 
If the $\mathcal{Z}_4$ is spontaneously broken, $\langle\phi\rangle\neq0$, the Dirac neutrino mass appears at tree level, and the particles must reside at a very large scale, without fine-tuning the Yukawa couplings. A spontaneously broken $\mathcal{Z}_4$ symmetry also introduces stable domain walls and destabilizes the DM candidate.} After electroweak symmetry breaking (EWSB), the Standard Model Higgs acquires a vacuum expectation value (VEV). In the presence of this soft term, the masses of the real ($\phi^R_{i}$) and imaginary ($\phi^I_{i}$) components of scalar $\phi_i$ are given by
\begin{eqnarray}
    M_{\phi^R_{i}}^2 &=& M_{\phi_i}^2 + \mu_{\phi_i}^2 + \frac{1}{2}\lambda_{\phi_i H} v^2, \\
    M_{\phi^I_{i}}^2 &=& M_{\phi_i}^2 - \mu_{\phi_i}^2 + \frac{1}{2}\lambda_{\phi_i H} v^2.
\end{eqnarray}
The corresponding mass-squared splitting is $
\Delta M_{\phi_i}^2 = M_{\phi^R_{i}}^2 - M_{\phi^I_{i}}^2 = 2\mu_{\phi_i}^2.$ We will demonstrate in Sec.~\ref{subsec:diracnumass} that this mass splitting plays a crucial role in generating Dirac neutrino masses at the one-loop level. Furthermore, the Higgs VEV ($v$) induces mixing between the singlet fermion and the neutral component of the doublet fermion through the interaction $\bar{\Psi}\tilde{H}\chi$, leading to singlet--doublet Dirac DM. Due to the mixing between $\chi$ and $\psi^0$ defined by $\sin\theta$, we get two mass eigenstates $\chi_1$ and $\chi_2$ with masses $M_{\chi_1}$ and $M_{\chi_2}$. We assume that $\chi_1$ is the lightest stable particle with mass $M_{\chi_1}\equiv M_{\rm{DM}}$, making it a natural candidate for the singlet-doublet Dirac DM. The details are provided in Appendix~\ref{app:dirac}. Similar to the Majorana case, we define two dark sectors: (a) sector 1, containing $\chi_1$, and (b) sector 2, comprising $\chi_2$, $\psi^\pm$, and other additional particles ($\phi_i$), while all SM particles are assigned to sector 0. We define the comoving number densities of sector 1 and sector 2 particles as $Y_1 \equiv n_{\chi_1}/s$ and $Y_2 \equiv (n_{\chi_2} + n_{\psi^\pm} +n_{\phi_{1,2,3}} )/s$, respectively. The coupled Boltzmann equations governing their evolution are the same as Eqs.~\ref{eq:Y1} and \ref{eq:Y2}. Finally, the CP-violating, out-of-equilibrium decay of the scalar fields generates a net lepton asymmetry via the Dirac leptogenesis mechanism, as will be discussed in Sec.~\ref{sec:leptoD}.
\subsection{Neutrino mass}\label{subsec:diracnumass}

\begin{figure}[H]
    \centering
    \includegraphics[scale=0.8]{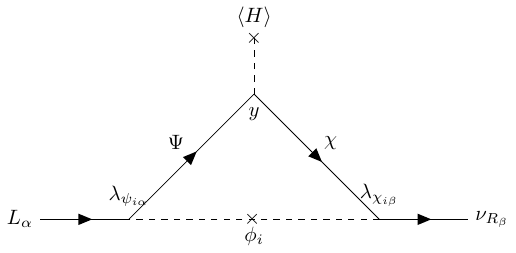}
    \caption{Realization of radiative Dirac neutrino mass at one loop level using the singlet-doublet Dirac DM.}
    \label{fig:neutrinomassDirac}
\end{figure}

As discussed in the previous section, the $Z_4$ symmetry is explicitly broken by the soft term $ \frac{1}{2}\mu_{\phi_i}^2 \left(\phi_i^2 + (\phi_i^\dagger)^2 \right).
$ As a result, the Dirac neutrino mass operator $\bar{L}\tilde{H}\nu_R$ is generated at the one-loop level \cite{Borah:2023dhk}. This arises from loop diagrams involving the singlet--doublet fermions $(\psi^0, \chi)$ and the three singlet scalar fields $\phi_i$, as illustrated in figure~\ref{fig:neutrinomassDirac}.
The one-loop neutrino mass can be calculated as:
\begin{eqnarray}
    \label{eq:neutrinomass}
    m_{\nu_{\alpha \beta}} =\frac{ \mu_{\phi}^2}{4 \pi^2}\sum_i (\lambda_{\Psi_{i\alpha}})^T\left( (\Delta M \sin{2\theta}) F(M_{\chi_1},M_{\chi_2},M_{\phi_1},M_{\phi_2})_{i}I_{ii} \right)(\lambda_{\chi_{i\beta}})
\end{eqnarray}
where $i$ represents the generation index of the scalar $\phi_i$, $\alpha$ and $\beta$ represent lepton flavor indices, $I_{ii}$ is the $3\times3$ identity matrix and $F(M_{\chi_1},M_{\chi_2},M_{\phi_1},M_{\phi_2})_{i} $ is the loop factor and is given by:
\begin{eqnarray}
F(M_{\chi_1},M_{\chi_2},M_{\phi_1},M_{\phi_2})_i&=&\frac{(M_{\chi_1}+M_{\chi_2})M_{\chi_2}^3 \ln{\left( \frac{M_{\chi_2}^2}{M_{\chi_1}^2} \right)}}{(M_{\chi_1}^2-M_{\chi_2}^2)(M_{\phi_{i_2}}^2-M_{\chi_2}^2)(M_{\phi_{i_1}}^2-M_{\chi_2}^2)} +\nonumber\\&&\frac{1}{(M_{\phi_{i_1}}^2-M_{\phi_{i_2}}^2)}\left( \frac{M_{\phi_{i_2}}^2(M_{\phi_{i_2}}^2+M_{\chi_1}M_{\chi_2})\ln{\left( \frac{M_{\phi_{i_2}}^2}{M_{\chi_1}^2} \right)}}{(M_{\phi_{i_2}}^2-M_{\chi_1}^2)(M_{\phi_{i_2}}^2-M_{\chi_2}^2)} \right.\nonumber\\ &&\left.-\frac{M_{\phi_{i_1}}^2(M_{\phi_{i_1}}^2+M_{\chi_1}M_{\chi_2})\ln{\left( \frac{M_{\phi_{i_1}}^2}{M_{\chi_1}^2} \right)}}{(M_{\phi_{i_1}}^2-M_{\chi_1}^2)(M_{\phi_{i_1}}^2-M_{\chi_2}^2)} \right)\nonumber\\    \label{eq:massfactor}
\end{eqnarray}

From Eq.~\ref{eq:neutrinomass}, we can define a biunitary transformation to diagonalize the neutrino mass matrix as: 
\begin{eqnarray}
    \label{eq:diagnumassdirac}
    D_{m_{ii}} = (V_{L_{\alpha i}})^\dagger m_{\nu_{\alpha \beta}} V_{R_{\beta i}},
\end{eqnarray}

where $V_L$ is the $U_{\rm PMNS}$ matrix and $V_R$ is some general unitary matrix (which has a similar form as the $U_{\rm PMNS}$) given as:
\begin{align}
V_R &= R_{23} \, R_{13} \, R_{12}.
\end{align}

In the above Eq.~the rotation matrices $R$ are given by

\begin{align}
R_{12} =
\begin{pmatrix}
\cos\theta_{1} & \sin\theta_{1} & 0 \\
-\sin\theta_{1} & \cos\theta_{1} & 0 \\
0 & 0 & 1
\end{pmatrix},\, \, 
R_{13} =
\begin{pmatrix}
\cos\theta_{3} & 0 & \sin\theta_{3} e^{-\iota\delta} \\
0 & 1 & 0 \\
-\sin\theta_{3} e^{\iota\delta} & 0 & \cos\theta_{3}
\end{pmatrix},\,\,
R_{23} =
\begin{pmatrix}
1 & 0 & 0 \\
0 & \cos\theta_{2} & \sin\theta_{2} \\
0 & -\sin\theta_{2} & \cos\theta_{2}
\end{pmatrix}.
\end{align}

Now we can further write the Eq.~\ref{eq:diagnumassdirac} as:
\begin{eqnarray}
    \label{eq:diracparameterization}
    D_{m_{ii}}=(V_{L_{\alpha i}})^\dagger(\lambda_{\psi_{k\alpha}})^T\Lambda_k I_{kk} \lambda_{\chi_{k\beta}}V_{R_{\beta i}}
\end{eqnarray}
where,
\begin{eqnarray}
    \Lambda_k = \frac{ \mu_{\phi_k}^2}{4 \pi^2}\left( (\Delta M \sin{2\theta}) F(M_{\chi_1},M_{\chi_2},M_{\phi_{k_1}},M_{\phi_{k_2}}) \right)
\end{eqnarray}
and $I_{kk}$ is a $3\times3$ identity matrix. Now using Eq.~\ref{eq:diracparameterization}, we can parameterize the couplings $\lambda_{\chi_{i\alpha}}$ and $\lambda_{\psi_{i\alpha}}$ as below:

\begin{eqnarray}           
\lambda_{\psi}=\sqrt{\Lambda^{-1}}^T\,R^T\, D_{\sqrt{m}}^T\,V_L^T,~~~~~~~~
\lambda_{\chi}=\sqrt{\Lambda^{-1}}\,R^{-1}\, D_{\sqrt{m}}\,V_R^{\dagger},\label{eq:ci}
\end{eqnarray}
where $R$ is a general $3\times3$ matrix.

\subsection{Muon anomalous magnetic moment}
\label{subsec:muong-2Dirac}

\begin{figure}[h]
    \centering
    \includegraphics[scale=1.0]{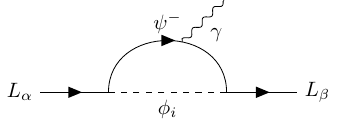}
    \caption{The Feynman diagram giving rise $(g-2)_\mu$ and charged lepton flavor violation.}
    \label{fig:lfvDirac}
\end{figure}

In our setup, the new positive contribution to the muon $(g-2)$ arises from the one-loop diagram involving the charged doublet fermion $\psi^{-}$ and the singlet scalars $\phi_i$s running in the loop, as shown in figure~\ref{fig:lfvDirac}.
This contribution to $(g-2)$ can be estimated in the same way as given in Eq. \ref{eq:g-2Majorana}, with the replacement $m_{\psi_i^-}\rightarrow m_{\psi^-}$ and $m_\phi\rightarrow m_{\phi_i}$.

\subsection{Charged lepton flavor violation}
\label{subsec:lfvDirac}

Similar to the Majorana case, in the present case, the additional particles ($\Psi$ and $\phi_i$) give rise to charged lepton flavor-violating processes such as $\mu\rightarrow e\gamma$ as shown in figure~\ref{fig:lfvDirac}. The branching ratio for the process $\mu\rightarrow e\gamma$ is calculated in the same way as given in Eq. \ref{eq:LFVMajorana}, with the replacement $m_{\psi_i^-}\rightarrow m_{\psi^-}$ and $m_\phi\rightarrow m_{\phi_i}$.

\begin{figure}[h]
    \centering \includegraphics[width=0.5\textwidth]{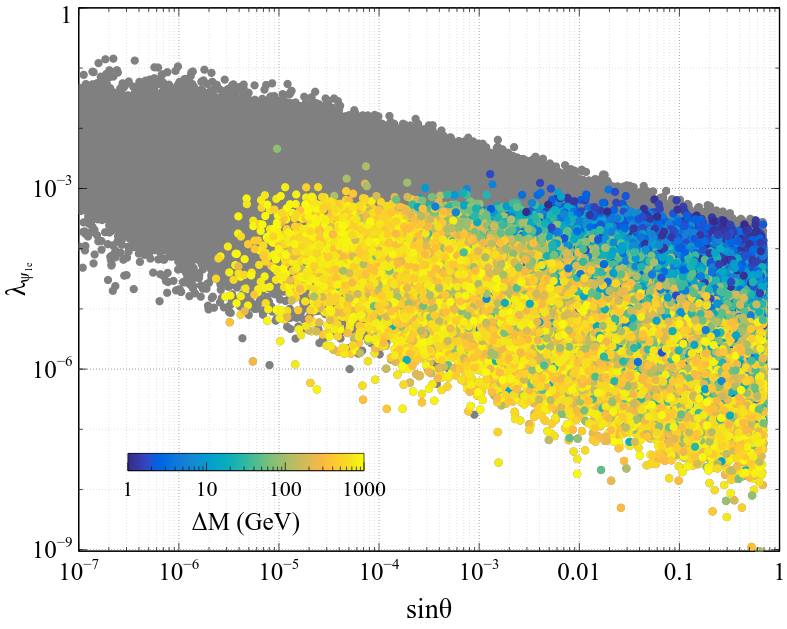}
    \caption{Parameter space consistent with neutrino mass, muon ($g-2$), and cLFV is shown with colored points in the plane of $\lambda_{\psi_{1e}}$ vs singlet-doublet mixing $\sin\theta$. The color code depicts the value of the singlet-doublet mass splitting $\Delta{M}$. The gray points in the background satisfy only the neutrino mass.}
\label{fig:couplingvsangledirac}
\end{figure}

\begin{table}[h!]
    \centering
    \begin{tabular}{|c|c|}
    \hline
    Parameter & Scan Range \\
    \hline
    $\log(M_{\chi_1}/\mathrm{GeV})$ & $[0,3]$ \\
    \hline
    $\log((M_{\chi_2} - M_{\chi_1})/\mathrm{GeV})$ & $[0,3]$ \\
    \hline
    $\log(\sin\theta)$ & $[-5,-0.15]$ \\
    \hline
    $\log((M_{\phi_1} - M_{\chi_2})/\mathrm{GeV})$ & $[0,3]$ \\
    \hline
    $M_{\phi_2}/M_{\phi_1},~~M_{\phi_3}/M_{\phi_2}$ & $[0,3]$ \\
    \hline
    $\mu_1/M_{\phi_1},~\mu_2/M_{\phi_2},~\mu_3/M_{\phi_3}$ & $10^{-2}$ \\
    \hline
    $\log(\theta_{12}),~\log(\theta_{13}),~\log(\theta_{23})$ & $[-10,4]$ \\
    \hline
    $\log(\delta)$ & $[-8,4]$ \\
    \hline
    $\log(R_{11})$ & $[-3,1]$ \\
    \hline
    $\log(R_{12})$ & $[-8,0]$ \\
    \hline
    $\log(R_{13}),~\log(R_{23}),~\log(R_{33})$ & $[-8,6]$ \\
    \hline
    $\log(R_{21})$ & $[-4,-1]$ \\
    \hline
    $\log(R_{22})$ & $[-4,1]$ \\
    \hline
    $\log(R_{31})$ & $[-5,-1]$ \\
    \hline
    $\log(R_{32})$ & $[-3,0]$ \\
    \hline
    \end{tabular}
\caption{Ranges in which the parameters are varied in the Dirac case. $R_{ij}$ are the elements of the $R$ matrix.}
\label{tab:dirac_scan}
\end{table}

In figure \ref{fig:couplingvsangledirac}, we show the parameter space consistent with neutrino mass generation, the muon anomalous magnetic moment $(g-2)\mu$, and cLFV by colored points in the $\lambda_{\psi_{1e}}$–$\sin\theta$ plane. The color scale indicates the singlet–doublet mass splitting $\Delta M$. Gray points in the background satisfy only the neutrino mass constraint. Similar to the Majorana case, we also observe that the minimum allowed value of $\lambda_{\psi_{1e}}$ decreases with increasing $\sin\theta$, reflecting the seesaw structure underlying neutrino mass generation. It is worth noting that the maximum allowed value of $\lambda_{\psi_{1e}}$ is $\sim \mathcal{O}(10^{-3})$, while the minimum value of $\sin\theta$ permitted by all constraints is $\sim \mathcal{O}(10^{-6})$. We vary the free parameters of the singlet-doublet Dirac DM model in the range as listed in table \ref{tab:dirac_scan}.

The mass-squared splitting $\mu_{\phi_i}$ is taken to be 1\% of the mass of $\phi_i$. The reason for this choice is as follows. The Dirac neutrino mass is $\propto \mu_\phi^2 \lambda_\psi \lambda_\chi\sin2\theta\times{\rm loop~factor}$. The loop factor is very small, which needs to be compensated for by the other couplings and the mixing angle. However, the left-side coupling $\lambda_\psi$ cannot be very large due to constraints from cLFV and $(g-2)_\mu$. A very large singlet–doublet mixing angle is also ruled out by direct detection constraints, which will be discussed later. Now, choosing a very small value of $\mu_\phi$ would require a large $\lambda_\chi$ to reproduce the observed neutrino mass. A large $\lambda_\chi$ will lead to strong washout effects in leptogenesis, and leptogenesis may not be viable unless $M_{\phi}$ is very large. Thus, we choose $\mu_\phi$ to be 1\% of $M_\phi$, which allows for relatively small values of $\lambda_\chi$ and enables successful leptogenesis at the TeV scale. For this reason, we cannot go below the TeV scale for successful leptogenesis, unlike in the Majorana case. To allow for leptogenesis below the TeV scale, $\mu_\phi$ would need to be increased further. We'll come back to this point while discussing Dirac leptogenesis from $\phi$ decay in Sec.~\ref{sec:leptoD}.

\subsection{Dark matter phenomenology}\label{subsec:DM_B}

\subsubsection{Thermal relic of dark matter}\label{subsec:relic_B}

\begin{figure}[h]
    \centering
 \includegraphics[width=0.475\textwidth]{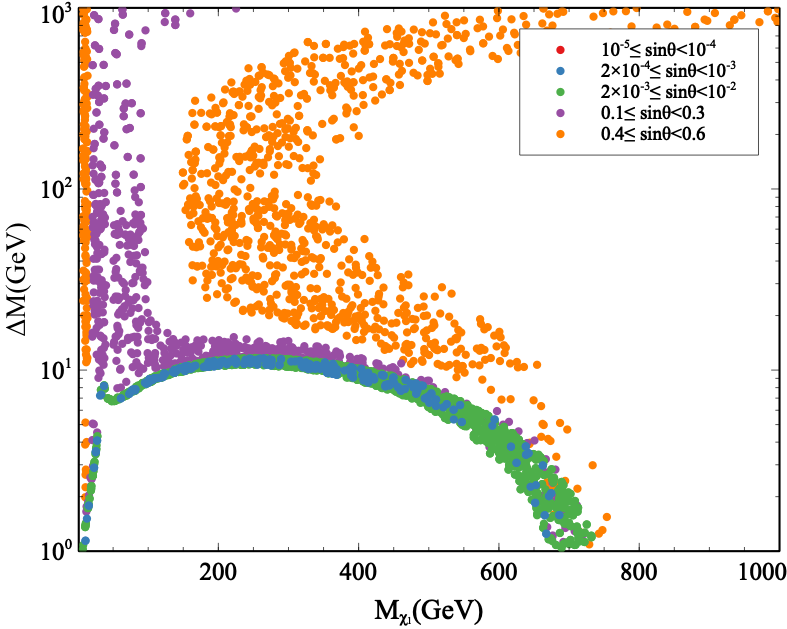}
\includegraphics[width=0.475\textwidth]{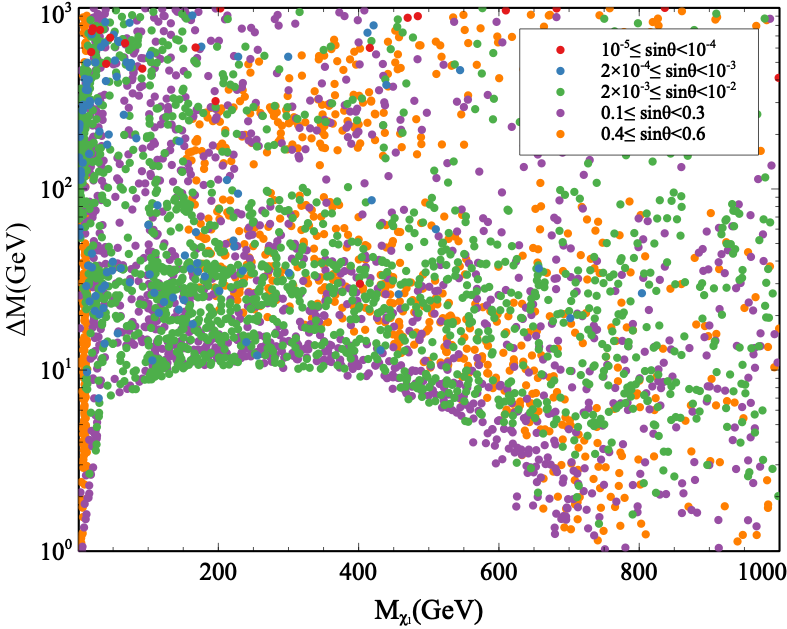}
    \caption{[\textit{Left}]: points satisfying correct relic shown in the plane of $\Delta{M}-M_{\rm{DM}}$ for the decoupling limit. [\textit{Right}]: same as left, but for a more general case. Note that all these points satisfy the constraints from neutrino mass, $(g-2)_{\mu}$, and cLFV.}  \label{fig:correctrelicDiracDM}
\end{figure}

As explained before, after the electroweak symmetry breaking, the neutral component of the doublet fermion mixes with the singlet fermion, giving rise to two Dirac mass eigenstates $\chi_1$ and $\chi_2$. The lightest of these states ($\chi_1$ as shown in Appendix~\ref{app:dirac}) becomes the DM candidate. We use \texttt{micrOMEGAs} to compute the relic density of DM. In this analysis, we take into account all the processes: annihilation, co-annihilation, and the conversion-driven processes. The parameters relevant for the relic calculation are the DM mass $M_{\rm DM}$, the singlet-doublet mass splitting $\Delta{M}$, the singlet-doublet mixing angle $\sin\theta$, singlet scalar masses, and the Yukawa couplings $\lambda_\psi,\lambda_\chi$. We vary these parameters in the model, keeping a mass hierarchy $M_{\chi_1}\leq M_{\chi_2}\leq M_{\phi_1}\leq M_{\phi_2}\leq M_{\phi_3}$ to allow successful Dirac leptogenesis, as will be discussed later. The Yukawa couplings are computed using Eq.~\ref{eq:ci}. In figure~\ref{fig:correctrelicDiracDM}, we present the regions of parameter space consistent with the observed relic density in the $M_{\rm{DM}}-\Delta{M}$ plane. These points also satisfy neutrino mass, $(g-2)_\mu$, and cLFV. The color code represents the value of the mixing angle $\sin\theta$. Here, we also consider two cases. We first studied a case where the three scalars are much heavier than the singlet–doublet fermion mass (left panel of figure~\ref{fig:correctrelicDiracDM}), which is the decoupling limit and gives rise to a similar parameter space as in the case of pure singlet-doublet Dirac DM \cite{Paul:2024prs,Bhattacharya:2018fus}. In the second case, we considered a more general mass spectrum, where the other dark sector particles can have masses close to the DM (right panel of figure~\ref{fig:correctrelicDiracDM}).

In the left panel of figure~\ref{fig:correctrelicDiracDM}, we present the correct DM relic satisfying points for the decoupling limit. The dependence of the parameters on the relic parameter space remains almost the same as the previously discussed Majorana case. As the DM mass increases, the annihilation cross-section drops, resulting in an overabundant relic of DM. The annihilation cross-section can be increased by increasing the $\Delta{M}$ (which increases the Yukawa coupling $y$) and the correct relic can be obtained. With an increase in $\Delta{M}$, $\sin\theta$ can be decreased accordingly to get the correct DM relic with $y$ being in the same order. For the smaller $\Delta{M}$, annihilation becomes subdominant, and the relic is mostly decided by the co-annihilation and conversion-driven processes. As the DM mass increases, $\Delta{M}$ keeps decreasing by maintaining an appropriate cross-section needed for the correct relic of DM. These features are clearly visible in the left panel of figure~\ref{fig:correctrelicDiracDM}.

In the right panel of figure~\ref{fig:correctrelicDiracDM}, we show the points satisfying correct relic in the same plane where the other dark sector states ($\phi_1$, $\phi_2$, and $\phi_3$) can be close to DM as well. It is interesting to note that the correct DM relic parameter space gets modified as compared to the pure singlet-doublet Dirac DM scenario. As discussed in Secs.~\ref{subsec:diracnumass}, \ref{subsec:muong-2Dirac}, and \ref{subsec:lfvDirac}, satisfying the $(g-2)_\mu$ and cLFV constraints requires the coupling $\bar{L}\Psi\phi$, denoted by $\lambda_{\psi}$, to be small $(\lambda_{\psi} \lesssim \mathcal{O}(10^{-3}))$. On the other hand, in order to reproduce the observed neutrino masses from low-energy data, the right-handed neutrino coupling $\bar{\nu}_R \chi \phi$, denoted by $\lambda_{\chi}$, needs to be large, leading to a seesaw-like behavior between the two couplings. Due to the large value of $\lambda_{\chi}$, annihilation processes mediated by $\phi$, such as $\chi_1 \chi_1 \rightarrow \nu_R \nu_R$ and $\chi_1 \bar{\chi}_1 \rightarrow \nu_R \bar{\nu}_R$, dominate and play a crucial role in determining the DM relic density. These processes become dominant when $\lambda_{\chi} > y$. However, for large mixing angles $\sin\theta$, the relic density is instead governed by singlet-doublet annihilation and co-annihilation processes, rather than the $\phi$-mediated channels.

\begin{figure}[h]
    \centering
    \includegraphics[width=0.475\textwidth]{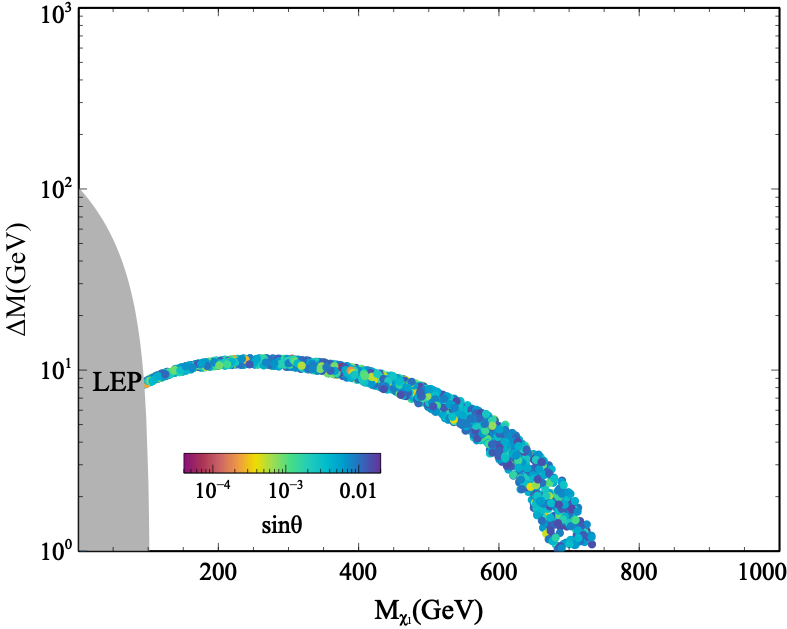}
    \includegraphics[width=0.475\textwidth]{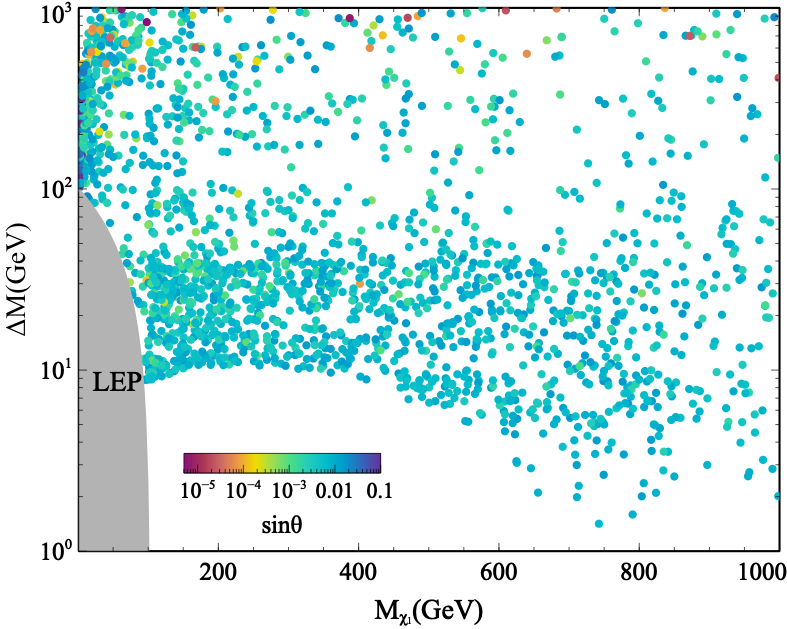}
    \caption{[\textit{Left}]: points satisfying correct relic and direct detection from the LZ and PANDAX-4T experiment are shown in the $\Delta M -M_{\rm DM}$ plane for the decoupling limit. The color code represents $\rm{sin}\theta$. [\textit{Right}]: same as the \textit{left} but for more general case. Note that all these points satisfy the constraints from neutrino mass, $(g-2)$, and cLFV.}
    \label{fig:crctrelafterDDDirac}
\end{figure}

\subsubsection{Direct detection prospects}\label{subsec:DD_B}

In this scenario, spin-independent (SI) DM direct detection is possible with the SM Higgs and $Z$ Boson exchange diagrams. We calculate the SI direct detection cross sections for the points shown in figure~\ref{fig:correctrelicDiracDM} and impose constraints from LZ \cite{LZ:2024zvo} and PANDAX-4T \cite{PandaX:2024qfu}. The allowed points from direct detection are shown in the plane of $M_{\rm{DM}}$ and $\Delta{M}$ in figure~\ref{fig:crctrelafterDDDirac}.
The gray shaded region represents the area excluded by LEP constraints on the doublet fermion. The \textit{left} plot corresponds to the decoupling limit (the scalar masses $M_{\phi_i}$ are much larger than the mass of singlet-doublet fermions), and the \textit{right} one corresponds to the general case, where the masses and mass splittings are varied according to Table~\ref{tab:dirac_scan}. The $\sin\theta$ allowed in the decoupling case is in the range of $\mathcal{O}(10^{-5})$ to $0.02$. On the other hand, the allowed $\sin\theta$ values in the general case can go up to $\mathcal{O}(10^{-2})$ for $M_{\rm{DM}}\gtrsim10 ~\rm{GeV}$. In the low mass region ($10~\rm{GeV}\gtrsim M_{\rm{DM}}\gtrsim1~\rm{GeV}$) LZ and PANDAX-4T do not give strong constraints on $M_{\rm{DM}}$. In this case, $\sin\theta$ can go up to $0.18$. 

\subsection{Dirac leptogenesis from $\phi$ decay}\label{sec:leptoD}

\begin{figure}[H]
    \centering
    \includegraphics[scale=0.6]{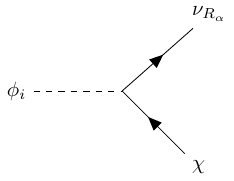}
    \includegraphics[scale=0.6]{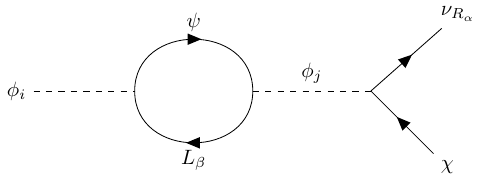}
    \includegraphics[scale=0.6]{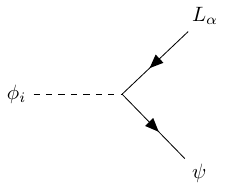}
    \includegraphics[scale=0.6]{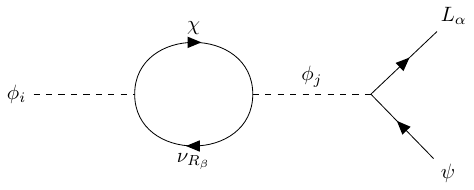}
    \caption{Tree level and one loop diagrams contributing to non-zero \textit{CP} asymmetry.}
    \label{fig:DiracCP}
\end{figure}

The baryon asymmetry can be obtained in this framework via Dirac leptogenesis \cite{Dick:1999je, Murayama:2002je, Boz:2004ga,Thomas:2005rs,Gu:2006dc,Bechinger:2009qk}. Since total lepton number is conserved, the \textit{CP}-violating out-of-equilibrium decays of $\phi_i$ produce equal and opposite lepton asymmetries in the left- and right-handed sectors through the channels $\phi_i \to L\Psi$ and $\phi_i \to \nu_R\chi$, respectively. The asymmetry stored in the left-handed sector survives and is partially converted into a baryon asymmetry through electroweak sphaleron processes. As there is no lepton number violation, the \textit{CP} asymmetry generated in the left and right-handed sectors is equal and opposite, \textit{$\epsilon_{L_i}=-\epsilon_{R_i}$}. The \textit{CP} asymmetry from figure \ref{fig:DiracCP}, is calculated to be \cite{Dick:1999je,Liu:1993tg}
\begin{eqnarray}
\label{eq:DiracCPasymmetry}
    \epsilon_{R_i}=\frac{1}{8\pi} \sum_{j\neq i}\frac{Im((\lambda_{\chi}^\dagger \lambda_{\chi})_{ij}(\lambda_{\psi}^\dagger\lambda_{\psi})_{ji})}{(\lambda_{\chi}^\dagger \lambda_{\chi})_{ii}+(\lambda_{\psi}^\dagger\lambda_{\psi})_{ii}}\frac{M_i^2}{M_i^2-M_j^2}=-\epsilon_{L_i}.
\end{eqnarray}

The evolution of the $\phi_i$ abundance and the associated left- and right-handed asymmetries ($Y_{\Delta_L}$ and $Y_{\Delta_{\nu_R}}$) in the early Universe is described by the following set of coupled Boltzmann equations (BEs)
\begin{eqnarray}
    \frac{d Y_{\phi_i}}{dz}&=& -\frac{1}{\mathcal{H}z} \left( (\Gamma_{\phi_i\rightarrow\nu_R \chi}(z)+\Gamma_{\phi_i\rightarrow L \Psi}(z)) (Y_{\phi_i}-Y_{\phi_i}^{eq}) \right)\\
    \frac{d Y_{\Delta_{\nu_R}}}{dz}&=& \frac{1}{\mathcal{H}z}\sum_i \left( \epsilon_{R_i} \Gamma_{\phi_i \rightarrow\nu_R \chi }(z)(Y_{\phi_i}-Y_{\phi_i}^{eq}) -\frac{1}{2}\frac{Y_{\phi_i}^{eq}}{Y_{\nu_R}^{eq}}\Gamma_{\phi_i \rightarrow\nu_R \chi}(z) Y_{\Delta_{\nu_R}} \right) \nonumber\\&&+ \frac{1}{\mathcal{H}z}n_\gamma \left<\sigma v\right>_{L\bar{\Psi}\rightarrow\nu_R \chi} \left( Y_{\Delta_{L}} - \frac{Y_{L}^{eq}}{Y_{\nu_R}}Y_{\Delta_{\nu_R}} \right)     Y_{\Psi}^{eq}\\
    \frac{d Y_{\Delta_{L}}}{dz}&=& \frac{1}{\mathcal{H}z} \sum_i \left( \epsilon_{L_i} \Gamma_{\phi_i \rightarrow L \Psi }(z)(Y_{\phi_i}-Y_{\phi_i}^{eq}) -\frac{1}{2}\frac{Y_{\phi_i}^{eq}}{Y_{L}^{eq}}\Gamma_{\phi_i \rightarrow L \Psi}(z) Y_{\Delta_{L}} \right)\nonumber\\&&+ \frac{1}{\mathcal{H}z}n_\gamma \left<\sigma v\right>_{\nu_R \chi \rightarrow L\bar{\Psi}} \left( Y_{\Delta_{\nu_R}} - \frac{Y_{\nu_R}^{eq}}{Y_{L}}Y_{\Delta_{L}} \right) Y_{\chi}^{eq}
    \label{eq:DiracBE}
\end{eqnarray}
where $z=M_{\phi_1}/T$, $\mathcal{H}$ is the Hubble parameter, $\Gamma$ and $\langle\sigma v\rangle$ are the the thermal averaged decay widths and cross-sections.
\begin{table*}[t]
    \centering
    \setlength{\tabcolsep}{6pt}
    \renewcommand{\arraystretch}{1}
    \begin{tabular}{|l|c|c|c|}
    \hline
    BP & DBP1 & DBP2\\
    \hline
    $M_{\chi_1}\,(\rm{GeV})$ & $204$ & $204$\\
    \hline
    $M_{\chi_2}\,(\rm{GeV})$ & $215$ & $215$\\
    \hline
    $\sin\theta$ & $0.01$ & $0.01$\\
    \hline
    $M_{\phi_1}\,(\rm{GeV})$ & $7.66\times10^3$ & $9.824\times10^3$\\
    \hline
    $M_{\phi_2}\,(\rm{GeV})$ & $7.66\times10^{6}$ & $9.824\times10^{6}$\\
    \hline
    $M_{\phi_3}\,(\rm{GeV})$ & $7.66\times10^7$ & $9.824\times10^7$\\
    \hline
    $\theta_1$ & $8.129\times 10^{-8}$ & $-3.779\times10^{-8}$\\
    \hline
    $\theta_2$ & $8.129\times 10^{-8}$ & $-3.779\times10^{-8}$\\
    \hline
    $\theta_3$ & $-8.141$ & $-1.252\times10^{-8}$\\
    \hline
    $\delta$ & $-2.799\times10^{-8}$ & $-1.893$\\
    \hline
    \end{tabular}
    \caption{Input parameters for the two benchmark points used in Dirac leptogenesis.}
\label{tab:Diractable}
\end{table*}
In this case the dilution factor is calculated to be $f$=30.05\footnote{$g_{*s}=106.75+\frac{7}{8}(2+8)+2=117.5$, is the relativistic d.o.f at the onset of leptogenesis and $g_{*0}$= 3.91 is the relativistic d.o.f today. The dilution factor f is then,$f=g_{*s}/g_{*0}=30.05$.}. The lepton to baryon asymmetry conversion factor in this case is $C=-0.33$\footnote{See appendix \ref{app:ltob_dir} for more details.}.

We solve the BEs for two benchmark points, DBP1 and DBP2, as listed in table \ref{tab:Diractable}, which satisfy the neutrino mass, $(g-2)_\mu$, cLFV, DM relic, and direct detection constraints. Here we take $\mu_1=1\%~{\rm of}~M_{\phi_1}$, $\mu_2=1\%~{\rm of}~M_{\phi_2}$, $\mu_3=1\%~{\rm of}~M_{\phi_3}$. We choose the rotation matrix to be
\begin{eqnarray}
R =
\begin{pmatrix}
-3.46\times10^{-1} + 1.31\times10^{-3} \iota 
& 8.72\times10^{1} - 2.23\times10^{-3} \iota 
& -4.56\times10^{2} + 8.73\times10^{2} \iota \\[6pt]

-1.72\times10^{-1} - 4.36\times10^{-1} \iota 
& -9.55\times10^{-3} - 3.88\times10^{-2} \iota 
& -3.44\times10^{-1} + 4.32\times10^{-2} \iota \\[6pt]

1.54\times10^{-1} - 3.00\times10^{-1} \iota 
& -5.52\times10^{-3} - 6.57\times10^{-2} \iota 
& -4.15\times10^{-1} - 2.08\times10^{-1} \iota
\end{pmatrix}.\nonumber\\
\end{eqnarray}
Following the parameterization in Eq. \ref{eq:ci}, the Yukawa coupling matrices come out to be

\begin{eqnarray}
    \lambda_{\chi} =
\begin{pmatrix}
-7.02\times10^{-11} - 7.63\times10^{-10}\iota &
-5.28\times10^{-3} + 1.56\times10^{-2}\iota &
-1.10\times10^{-2} + 1.49\times10^{-2}\iota \\[6pt]
-1.44\times10^{-9} + 9.37\times10^{-9}\iota &
4.40\times10^{-2} - 1.19\times10^{-1}\iota &
5.47\times10^{-2} - 4.77\times10^{-2}\iota \\[6pt]
-7.94\times10^{-10} + 2.71\times10^{-10}\iota &
1.11\times10^{-2} - 1.41\times10^{-3}\iota &
5.99\times10^{-3} + 2.34\times10^{-3}\iota
\end{pmatrix},\nonumber\\
\end{eqnarray}
and 
\begin{eqnarray}
    \lambda_{\Psi} =
\begin{pmatrix}
-3.35\times10^{-4} - 4.71\times10^{-4}\iota &
6.00\times10^{-4} - 3.09\times10^{-3}\iota &
1.47\times10^{-3} - 1.19\times10^{-3}\iota \\[6pt]
3.51\times10^{-5} - 9.38\times10^{-6}\iota &
-5.55\times10^{-5} - 5.24\times10^{-4}\iota &
-1.33\times10^{-5} - 3.71\times10^{-4}\iota \\[6pt]
-9.17\times10^{-5} + 3.19\times10^{-5}\iota &
-3.56\times10^{-3} - 1.27\times10^{-3}\iota &
-2.13\times10^{-3} - 1.55\times10^{-3}\iota
\end{pmatrix}.\nonumber\\
\end{eqnarray}

\begin{figure}[h]
    \centering
    \includegraphics[width=0.475\textwidth]{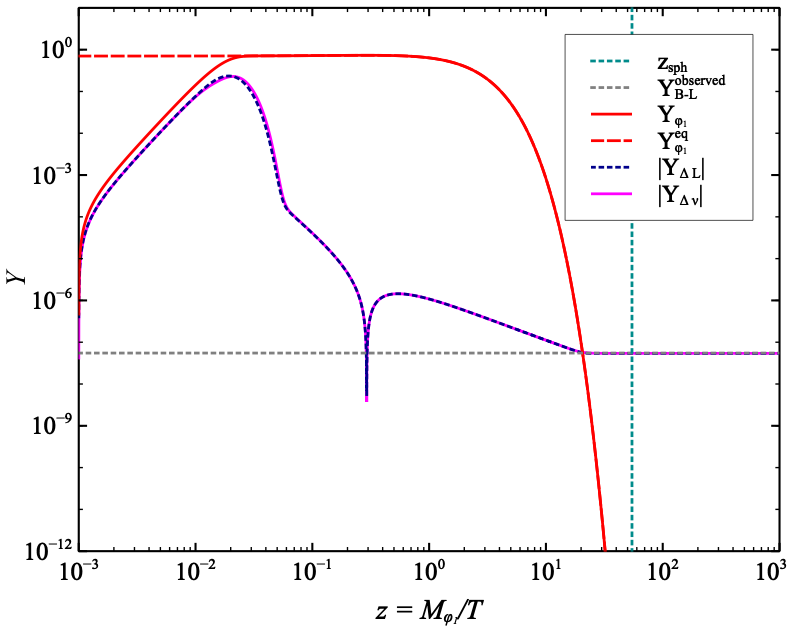}
    \includegraphics[width=0.475\textwidth]{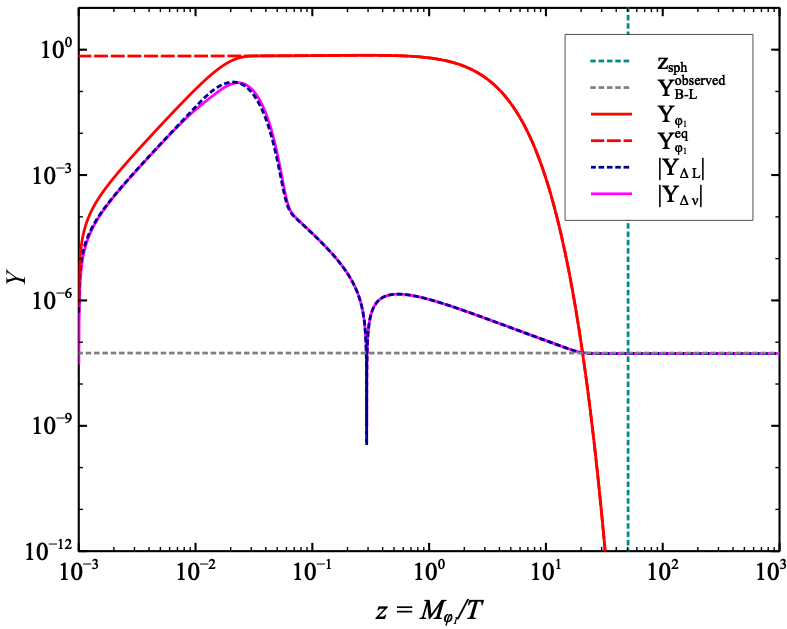}
    \caption{[\textit{Left}:] cosmological evolution of $\phi_1$ abundance, $\nu_R$ asymmetry, and lepton asymmetry are shown with red, magenta, and dotted blue colored lines, respectively, for DBP1. The equilibrium abundance of $\phi_1$ is shown with a red dashed line. The dark-cyan vertical dotted line represents the sphaleron transition temperature, $T_{\rm sph}\simeq131.7$ GeV \cite{DOnofrio:2014rug}. The gray dotted horizontal line corresponds to the correct baryon asymmetry value. [\textit{Right}:] the same as in the \textit{left} panel, but for DBP2.}
    \label{fig:DBP-1}
\end{figure}

This leads to decay parameters $K_{R_1} = 3.3624 \times 10^{-4}$ and $K_{L_1} = 1.2719 \times 10^{6}$. Although the decay parameter in the right-handed sector is very small, a strong washout occurs in both the left- and right-handed sectors due to the large value of the left-handed decay parameter. The \textit{CP} asymmetry parameter is calculated to be $\epsilon_{R_1}=-\epsilon_{L_1}=9.189\times 10^{-1}$. It is important to note that the \textit{CP} asymmetry due to $\phi_2$ and $\phi_3$ are small compared to $\phi_1$ ($\epsilon_{R_2}=4.297\times10^{-6}=-\epsilon_{L_2}$ and $\epsilon_{R_3}=2.868\times10^{-3}=-\epsilon_{L_3}$), while their corresponding decay parameters are very large ($K_{R_2}=6.06\times10^{7}$, $K_{L_2}=9.03\times10^{4}$, $K_{R_3}=2.08\times10^{6}$ and $K_{L_3}=3.93\times10^{3}$). Consequently, the contribution of $\phi_{2,3}$ to the production of the asymmetry is negligible. The primary effect of $\phi_{2,3}$ would be to wash out the asymmetry generated by $\phi_1$ due to their large decay parameters. However, since $M_{\phi_2} = 7.66 \times 10^{6}$ GeV and $M_{\phi_3} = 7.66 \times 10^{7}$ GeV, $\phi_2$ and $\phi_3$ decay well before $z \sim 10^{-5}$ and therefore does not significantly influence the evolution of $\phi_1$. For completeness, we nevertheless include $\phi_{2,3}$ in the Boltzmann equations, although it has no impact on the final asymmetry generated by $\phi_1$. In the \textit{left} panel of figure~\ref{fig:DBP-1}, we show the evolution of the asymmetries in the left- and right-handed sectors, along with the $\phi_1$ abundance, using different colored lines as indicated in the inset. The gray dotted line denotes the observed lepton asymmetry, while the dark cyan dotted line marks the sphaleron decoupling temperature. This benchmark point lies in the strong washout regime, leading to a significant suppression of the final asymmetry. Nevertheless, due to the large \textit{CP} asymmetry parameters, the final asymmetry settles within the observed range. For this benchmark point, the lepton asymmetry is found to be $Y_{\Delta L} = 5.5 \times 10^{-8}$, which is subsequently converted into the baryon asymmetry via electroweak sphaleron processes, yielding $\eta_B \simeq 6 \times 10^{-10}$. 

In the \textit{right} panel of figure \ref{fig:DBP-1}, we show the evolution of the asymmetries and $\phi_1$ for the DBP2. We take the same rotation matrix as in DBP1. The resulting Yukawa couplings are given as
\begin{eqnarray}
\lambda_{\chi} =
\begin{pmatrix}
1.13\times10^{-9} - 3.07\times10^{-10}\iota &
-9.02\times10^{-3} + 9.80\times10^{-3}\iota &
8.16\times10^{-3} - 1.92\times10^{-2}\iota \\[6pt]
-6.77\times10^{-9} - 8.55\times10^{-11}\iota &
4.00\times10^{-2} - 1.21\times10^{-2}\iota &
-5.77\times10^{-2} + 1.27\times10^{-1}\iota \\[6pt]
-2.71\times10^{-10} - 5.35\times10^{-10}\iota &
2.60\times10^{-3} + 2.64\times10^{-3}\iota &
-1.24\times10^{-2} + 6.93\times10^{-4}\iota
\end{pmatrix},\nonumber\\
\end{eqnarray}
and
\begin{eqnarray}
    \lambda_{\Psi} =
\begin{pmatrix}
-3.34\times10^{-4} - 4.70\times10^{-4}\iota &
5.99\times10^{-4} - 3.09\times10^{-3}\iota &
1.46\times10^{-3} - 1.18\times10^{-3}\iota \\[6pt]
3.51\times10^{-5} - 9.38\times10^{-6}\iota &
-5.55\times10^{-5} - 5.24\times10^{-4}\iota &
-1.33\times10^{-5} - 3.71\times10^{-4}\iota \\[6pt]
-9.17\times10^{-5} + 3.19\times10^{-5}\iota &
-3.56\times10^{-3} - 1.27\times10^{-3}\iota &
-2.13\times10^{-3} - 1.55\times10^{-3}\iota
\end{pmatrix}.\nonumber\\
\end{eqnarray}
The decay parameters for this BP is given as $K_{R_1}=1.3691\times 10^{-4}$, and $K_{L_1}=9.8861\times 10^{5}$. The \textit{CP} asymmetries are calculated to be $\epsilon_{R_1}=-\epsilon_{L_1}=6.9804\times10^{-1}$. The CP asymmetries of the heavier scalars are $\epsilon_{R_2}=3.909\times10^{-4}=-\epsilon_{L_2}$ and $\epsilon_{R_3}=8.634\times10^{-3}=-\epsilon_{L_3}$. And the washout parameters of the heavier scalars are $K_{R_2}=5.58\times10^{6}$,$K_{L_2}=7.03\times10^{4}$,$K_{R_3}=5.80\times10^{6}$,$K_{L_3}=3.06\times10^{3}$. Similar to DBP1, the contributions of $\phi_{2,3}$ are negligible due to its extremely large decay parameter and large mass compared to $\phi_1$ which ensures that they decay well before it affects the asymmetry produced by $\phi_1$. This BP also corresponds to a strong washout regime, and we observe a substantial suppression of the final asymmetry. Thanks to the large \textit{CP} asymmetry parameters, the final asymmetry settles at the observed ballpark. The final lepton asymmetry is found to be $Y_{\Delta L}=5.5\times10^{-8}$, which translates to the baryon asymmetry as $\eta_B\simeq6.0\times10^{-10}$.
For these BPs, the DM relic is found to be 0.12, and the direct detection cross-section is $7.889\times10^{-49}~\rm cm^{-2}$, which is well within the sensitivity of DARWIN \cite{DARWIN:2016hyl}.

\section{Collider signatures: displaced vertex and prompt decay searches}\label{sec:coll}

\begin{figure}
    \centering
    \includegraphics[width=0.475\textwidth]{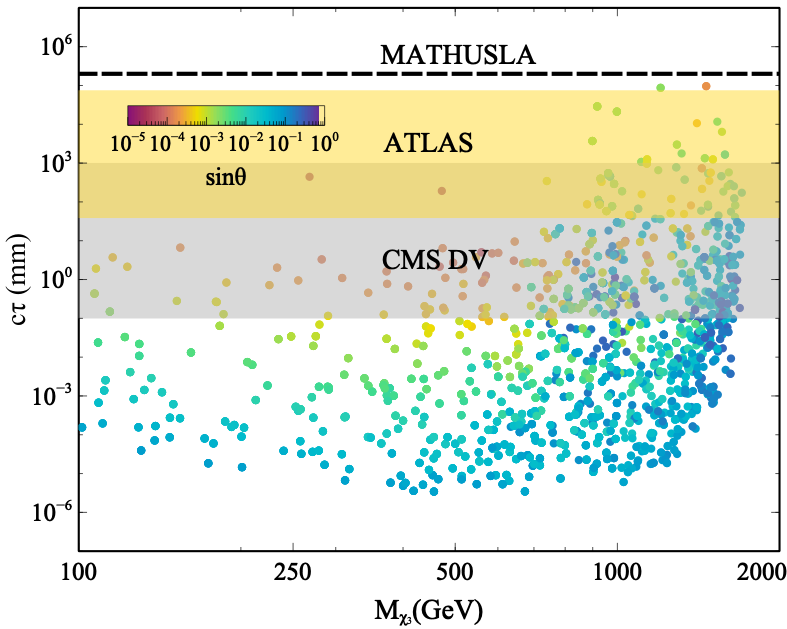}
    \includegraphics[width=0.475\textwidth]{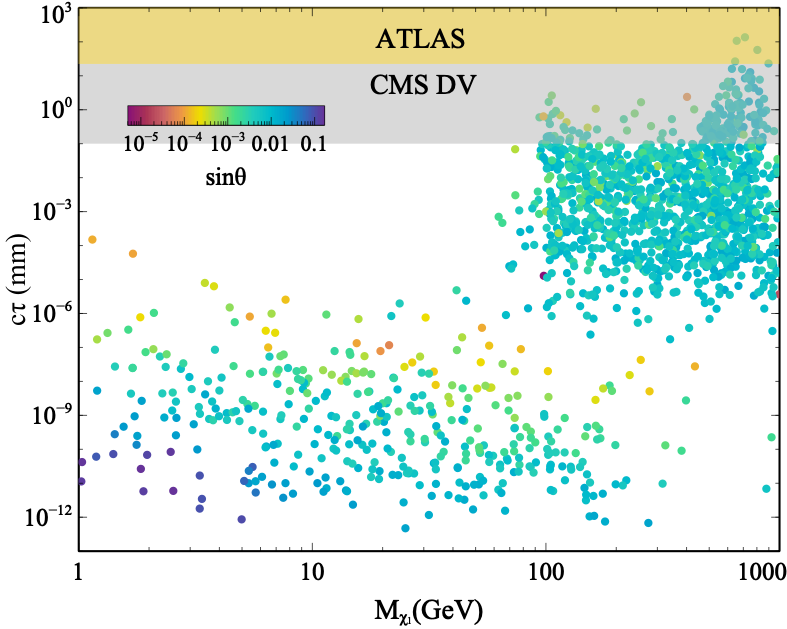}
    \caption{\textit{Left:} Decay length of the doublet fermion as a function of DM mass with $\sin\theta$ in the color code for the Majorana case. \textit{Right:} Decay length of the doublet fermion as a function of DM mass with $\sin\theta$ in the color code for the Dirac case.}
    \label{fig:displacedvertexsign}
\end{figure}

In this section, we briefly discuss the collider signatures of the model. The first generation of charged doublet fermion, being in the sub-TeV range, $\psi_1^\pm$ can be produced via gauge interactions and subsequently decay into DM and charged leptons through singlet–doublet mixing, mediated by off-shell gauge bosons. The collider signature depends on its decay length, which is controlled by the mixing angle $\sin\theta$. For small mixing \cite{Paul:2024prs,Paul:2025spm}, the decay width is suppressed, leading to long-lived charged particles that can give rise to displaced vertex signatures at the LHC \cite{ATLAS:2022gbw,CMS:2024trg} and at future detectors such as MATHUSLA \cite{MATHUSLA:2019qpy}. On the other hand, when the decay length is smaller than $10^{-4}$ m \cite{Paul:2024prs,Paul:2025spm}, the decay becomes prompt before reaching the detector volume, resulting in standard collider signatures with missing transverse energy and multilepton final states, such as $pp \rightarrow 3\ell + E_T^{\rm miss}$\cite{CMS:2021cox,ATLAS:2021moa}. These complementary signatures provide important probes of the model across different regions of parameter space \cite{Paul:2024prs,Paul:2025spm}. 
In figure~\ref{fig:displacedvertexsign}, we show the decay length of $\psi_1^\pm$ as a function
of DM mass. The color code represents $\sin\theta$.
These points in the figure~\ref{fig:displacedvertexsign} left and right correspond to the same points as in the right panel of figure~\ref{fig:crctrelafterDD} (for Majorana) and figure~\ref{fig:crctrelafterDDDirac} (for Dirac), respectively.
The sensitivity range of CMS is shown in the light gray shaded region, the sensitivity of ATLAS is shown by the light yellow shaded region, and the MATHUSLA projection is shown by the black dashed line.
We note that in figure~\ref{fig:displacedvertexsign} (left), many of the points lie within and above the sensitivity ranges of CMS and ATLAS; this is due to the small mass splitting between the first generation singlet-doublet $\Delta{M}$, but for larger mass splitting, many points fall below the CMS displaced vertex sensitivity range. Similarly, in figure~\ref{fig:displacedvertexsign} (right), many points lie within the sensitivity of CMS. This is also due to the smaller mass splitting $\Delta{M}$ between singlet-doublet fermions. These plots can be verified by comparing the mass splittings with figure~\ref{fig:crctrelafterDD} (right) and figure~\ref{fig:crctrelafterDDDirac} (right).

\section{Conclusion and future outlook}\label{sec:concl}

In this work, we explored the possibility of explaining the tiny neutrino mass and baryon asymmetry of the Universe via leptogenesis in well-motivated singlet-doublet dark matter (DM) scenarios. We considered two possibilities: a) singlet-doublet Majorana DM and b) singlet-doublet Dirac DM.

In the case a), the standard model is extended with three generations of singlet-doublet vector-like fermions ($\chi_i,\Psi_i$), and one singlet scalar $\phi$, all of which are odd under $\mathcal{Z}_2$, and belong to the ``dark" sector. The tiny Majorana neutrino mass is realized at one loop level, with dark sector particles running in the loop. The first generation of the singlet–doublet fermions gives rise to a Majorana DM candidate. In contrast, the \textit{CP}-violating, out-of-equilibrium decays of the second and third generation singlets, \textit{i.e.}, $N_2$ and $N_3$, into the corresponding doublets $\Psi_2$ and $\Psi_3$, generate a net asymmetry in the doublet sector. This asymmetry is subsequently transferred to the SM lepton sector via the interaction $\Psi\rightarrow L \phi$. The net lepton asymmetry is then converted to the baryon asymmetry via the electroweak sphalerons. Here, we showed that, in this case, correct baryon asymmetry can be achieved even in the sub-TeV range.

In the case b), the SM is extended by one generation of singlet–doublet vector-like fermions $(\chi,\Psi)$, three generations of singlet scalars $\phi_i$, and three generations of light right-handed counterparts of the SM neutrinos, $\nu_R$. All BSM fields, except the $\nu_R$, are odd under a $\mathcal{Z}_2$ symmetry and constitute the \textit{dark} sector. Since all fermions are of Dirac nature, the light neutrino masses are generated radiatively at one-loop level, with the dark sector particles running in the loop, similar to the previous case. After electroweak symmetry breaking, mixing between the singlet and the neutral component of the doublet fermion gives rise to singlet–doublet Dirac DM. The \textit{CP}-violating, out-of-equilibrium decays of the scalar fields generate equal and opposite asymmetries in the left- and right-handed sectors via the Dirac leptogenesis mechanism, as total lepton number is conserved in this setup. The observed baryon asymmetry in this case can be achieved at the TeV scales.

In both scenarios, we find that the particles relevant to the phenomenology lie in the GeV to TeV mass range, making the model testable at current and future collider experiments \cite{Paul:2024prs,Paul:2025spm,Dey:2025pcs}, as well as in DM direct detection searches. Since the mass splitting between the neutral and the charged components of the doublet $\Psi$ is small (typically of the order of $\mathcal{O}(10^2)~\rm{MeV}$), the charged component can give rise to a displaced vertex signature. The light right-handed neutrinos involved in Dirac leptogenesis thermalize in the early Universe and contribute to the relativistic degrees of freedom, yielding $\Delta N_{\rm eff} \sim 0.14$. This additional contribution lies within the projected sensitivity of upcoming experiments such as CMB-S4 \cite{Abazajian:2019eic} and CMB-HD \cite{CMB-HD:2022bsz}.

\acknowledgments
P.K.P. acknowledges the Ministry of Education, Government of India, for providing financial support for his research via the Prime Minister’s Research Fellowship (PMRF) scheme.

\appendix

\section{Singlet-doublet Majorana dark matter model}\label{app:majo}
Due to the unbroken $\mathcal{Z}_2$ symmetry combination of $N_1$ and $\Psi_1 = (\psi^0\quad \psi^-)^T\equiv(\psi_L^0+\psi_R^0 \quad \psi^-)^T$ can give rise to a singlet-doublet Majorana DM. The relevant DM Lagrangian reads as:
\begin{eqnarray}
    \mathcal{L}_{\rm DM} &=& \bar{\Psi}_1\iota\gamma^\mu \mathcal{D}_\mu \Psi_1 - M_{\Psi_1}\bar{\Psi}_1\Psi_1 + \bar{N}_1\iota\gamma^\mu \partial_\mu N_1 - \frac{1}{2}M_{N_1}\bar{N_1^c}N_1 \nonumber\\&&- \frac{Y_{11}}{\sqrt{2}}\bar{\Psi}_1\tilde{H}(N_1+N_1^c)  + {\rm H.c.}
\end{eqnarray}
The neutral fermion mass matrix can be written in the left-handed Weyl spinor basis 
$\begin{pmatrix}
    \psi_L^0&(\psi_R^0)^c&N_1^c
\end{pmatrix}^T$ as

\begin{eqnarray}
\label{massmatrix}
    \mathcal{M}=
    \begin{pmatrix}
     0 &M_\Psi& \frac{m_D}{\sqrt{2}}\\
     M_\Psi & 0 & \frac{m_D}{\sqrt{2}}\\
     \frac{m_D}{\sqrt{2}} & \frac{m_D}{\sqrt{2}} & M_{N_1}
\end{pmatrix},
\end{eqnarray}
where $m_D = y_{11} v/\sqrt{2}$. This mass matrix can be diagonalized using a unitary matrix $\mathcal{U}$ which is defined as:
\begin{equation}
    \begin{split}
        \mathcal{U} &=
    \begin{pmatrix}
        1 & 0 & 0 \\
        0 & e^{\iota\delta} & 0\\
        0 & 0 & 1
    \end{pmatrix}
    \begin{pmatrix}
        \cos \theta_1 & 0 & -\sin \theta_1 \\
        0 & 1 & 0\\
        \sin \theta_1  & 0  & \cos \theta_1 
    \end{pmatrix}
    \begin{pmatrix}
        1 & 0 & 0\\
        0 & \cos \theta_2  & -\sin \theta_2 \\
        0 & \sin \theta_2  & \cos \theta_2 \\
    \end{pmatrix}
    \begin{pmatrix}
        \cos \theta_3 & -\sin \theta_3 & 0 \\
        \sin \theta_3 & \cos \theta_3 & 0\\
        0 & 0 & 1
    \end{pmatrix}
\end{split}\label{eq:Rmatrix}
\end{equation}
Keeping in mind the texture of the mass matrix \ref{massmatrix}, we can use a simple version of the unitary matrix needed for diagonalization as $\mathcal{U}(\theta)=U_\delta(\delta=\pi/2)U_{13}(\theta_{1}=\theta).U_{23}(\theta_{2}=0).U_{12}(\theta_{3}=\pi/4)$, which is essentially defined by $\theta_{1}=\theta$:
\begin{eqnarray}
\mathcal{U}(\theta)=
    \begin{pmatrix}
        1&0&0\\
        0&e^{\iota\pi/2}&0\\
        0&0&1
    \end{pmatrix}
    \begin{pmatrix}
        \frac{1}{\sqrt{2}}\rm{cos}\,\theta & \frac{1}{\sqrt{2}}\rm{cos}\,\theta&\rm{sin}\,\theta\\
        -\frac{1}{\sqrt{2}}&\frac{1}{\sqrt{2}} &0\\
        -\frac{1}{\sqrt{2}}\rm{sin}\,\theta&-\frac{1}{\sqrt{2}}\rm{sin}\,\theta&\rm{cos}\,\theta 
    \end{pmatrix}
\end{eqnarray}
After using the relation $\mathcal{M}_{diag}=\mathcal{U}.\mathcal{M}.\mathcal{U}^T$, the mass matrix changes to:
\begin{eqnarray}
\label{diagmassmatrix}
    \mathcal{M}_D =
    \begin{pmatrix}
        M_{\Psi_1} \rm{cos}^2\theta + M_{N_1} \rm{sin}^2\theta + m_D \,\rm{sin}2\theta & 0 & m_D\,\rm{cos}\,2\theta + \frac{M_{N_1}-M_{\Psi_1}}{2}\rm{sin}\,2\theta\\
        0 & M_{\Psi_1} & 0\\
        m_D\,\rm{cos}\,2\theta + \frac{M_{N_1}-M_{\psi_1}}{2}\rm{sin}\,2\theta & 0 &
        M_{N_1} \rm{cos}^2\theta + M_{\Psi_1} \rm{sin}^2\theta - m_D \,\rm{sin}2\theta
    \end{pmatrix}
\end{eqnarray}
Now in order for \ref{diagmassmatrix} to be diagonal, the mixing angle $\theta$ has to be:
\begin{eqnarray}
    {\rm tan} 2\theta = \frac{2 \,m_D}{M_{\Psi_1}-M_{N_1}} = \frac{\sqrt{2}y_{11} v}{M_{\Psi_1}-M_{N_1}}.
\end{eqnarray}
Now the unphysical basis, $\begin{pmatrix}
    \psi_L^0&(\psi_R^0)^c&N_1^c
\end{pmatrix}^T$ is related to the physical mass basis, $\begin{pmatrix}
    \chi_1&\chi_2&\chi_3
\end{pmatrix}^T$ (where, $\chi_i =\frac{\chi_{iL}+(\chi_{iL})^c}{\sqrt{2}}$ with $i=1,2,3$) through the following unitary transformation:
\begin{eqnarray}
    \begin{pmatrix}
        \chi_{1L}\\ \chi_{2L}\\\chi_{3L}
    \end{pmatrix}=\mathcal{U} \begin{pmatrix}
    \psi_L^0\\(\psi_R^0)^c\\N_1^c
\end{pmatrix}
\end{eqnarray}
which gives the relations:
\begin{eqnarray}
    \chi_{1L}&=&\frac{1}{\sqrt{2}}\rm{cos}\,\theta(\psi_L^0 + (\psi_R^0)^c) + sin\,\theta\, (N_1)^c,\nonumber\\
    \chi_{2L}&=&-\frac{\iota}{\sqrt{2}}(\psi_L^0 - (\psi_R^0)^c),\\
    \chi_{3L}&=&-\frac{1}{\sqrt{2}}\rm{sin}\,\theta\,(\psi_L^0 + (\psi_R^0)^c) + \rm{cos} \,\theta \, (N_1)^c.\nonumber
\end{eqnarray}
Therefore, the corresponding mass eigenvalues of the physical states $\chi_1$, $\chi_2$ and $\chi_3$ are:
\begin{eqnarray}
    M_{\chi_1} &=& M_{\Psi_1} \cos^2\theta + M_{N_1} \sin^2\theta +\frac{y_{11} v}{\sqrt{2}}\rm{sin}2\theta\\
    M_{\chi_2} &=& M_{\Psi_1}\\
    M_{\chi_3} &=& M_{\Psi_1} \sin^2\theta + M_{N_1} \cos^2\theta -\frac{y_{11} v}{\sqrt{2}}\rm{sin}2\theta
\end{eqnarray}
Here, we have considered $y_{11}v\sqrt{2}\ll M_{N_1}<M_{\Psi_1}$. Hence, $M_{\chi_1}\gtrsim M_{\chi_2}>M_{\chi_3}$. Therefore, $\chi_3$ becomes the stable DM candidate. Now the Yukawa coupling $y_{11}$ can be expressed as:
\begin{eqnarray}
    y_{11}=\frac{\Delta M\,\rm{sin}2\theta}{\sqrt{2}v}
\end{eqnarray}
where, $\Delta M = M_{\chi_1}-M_{\chi_3}\approx M_{\chi_2}-M_{\chi_3}$.
The free parameters in the DM phenomenology are $\{ M_{\chi_3}, \Delta M, \,\rm{sin}\,\theta \}$. The singlet fermion, being the lightest among the dark sector states, serves as the DM candidate.
We define two dark sectors: (a) sector 1, containing $\chi_3$, and (b) sector 2, comprising $\chi_1,~\chi_2$, $\psi_1^\pm$, and other additional particles ($N_{2,3},\,\Psi_{2,3}$ and $\phi$), while all SM particles are assigned to sector 0. We define the comoving number densities of sector 1 and sector 2 particles as $Y_1 \equiv n_{\chi_3}/s$ and $Y_2=(n_{\chi_1}+n_{\chi_2}+n_{\psi_1^\pm}+n_{N_{2,3}}+n_{\Psi_{2,3}}+n_\phi)/s$, respectively. The coupled Boltzmann equations governing their evolution can be written as \cite{Alguero:2022inz,Paul:2024prs,Paul:2025spm}
\begin{eqnarray}
		\frac{dY_1}{dT} &=&   \frac{1}{3\mathcal{H}}\frac{ds}{dT} \left[    \langle \sigma_{1100} v \rangle ( Y_1^2 - {Y_1^{\rm eq}}^2) +    \langle \sigma_{1122} v \rangle \left( Y_1^2 - Y_2^2  \frac{{Y_1^{\rm eq}}^2}{{Y_2^{\rm eq}}^2}\right)  + \langle \sigma_{1200} v \rangle ( Y_1 Y_2 - Y_1^{\rm eq}Y_2^{\rm eq})\right. \nonumber\\
		&&+\left.  \langle \sigma_{1222} v \rangle \left( Y_1 Y_2 - Y_2^2   \frac{Y_1^{\rm eq}}{Y_2^{\rm eq}} \right) -\langle \sigma_{1211} v \rangle \left( Y_1 Y_2 - Y_1^2   \frac{Y_2^{\rm eq}}{Y_1^{\rm eq}} \right)
		-\frac{ \Gamma_{2\rightarrow 1}}{s}\left( Y_2 -Y_1 \frac{Y_2^{\rm eq}}{Y_1^{\rm eq}}  \right)        \right] ,\nonumber\\
		\label{eq:Y1}
	\end{eqnarray}
	\begin{eqnarray}
		\frac{dY_2}{dT} &=&   \frac{1}{3\mathcal{H}}\frac{ds}{dT}\left[    \langle \sigma_{2200} v \rangle ( Y_2^2 - {Y_2^{\rm eq}}^2) -    \langle \sigma_{1122} v \rangle \left( Y_1^2 - Y_2^2  \frac{{Y_1^{\rm eq}}^2}{{Y_2^{\rm eq}}^2}\right) +  \langle \sigma_{1200} v \rangle ( Y_1 Y_2 - Y_1^{\rm eq}Y_2^{\rm eq}) \right. \nonumber \\
		&&- \left. \langle \sigma_{1222} v \rangle \left( Y_1 Y_2 - Y_2^2   \frac{Y_1^{\rm eq}}{Y_2^{\rm eq}} \right)
		+\langle \sigma_{1211} v \rangle \left( Y_1 Y_2 - Y_1^2   \frac{Y_2^{\rm eq}}{Y_1^{\rm eq}} \right)  + \frac{ \Gamma_{2\rightarrow 1}}{s}\left( Y_2 -Y_1 \frac{Y_2^{\rm eq}}{Y_1^{\rm eq}}  \right)        \right],\nonumber\\
		\label{eq:Y2}
	\end{eqnarray}
	where $Y_i^{\rm eq}\left(=\frac{n_i^{\rm eq}}{s}\right)$ are the equilibrium abundances, $\mathcal{H}$ is the  Hubble parameter, the entropy density, $s=\frac{2\pi^2}{45}g_{*s}T^3$, $\langle \sigma_{\alpha\beta\gamma\delta} v\rangle$ are the thermally averaged cross-sections for processes involving annihilation of particles of sectors $\alpha\beta\rightarrow \gamma\delta$ and $\Gamma_{2\rightarrow 1}$ is the conversion term, which includes both the interaction rate of the co-scattering process and decay. The DM relic is then calculated to be 
    \begin{eqnarray}
        \Omega_{\rm DM}h^2=2.742\times10^8\times(M_{\chi_3}Y_1+M_{\chi_2}Y_2).
    \end{eqnarray}

\section{Singlet-doublet Dirac dark matter model}\label{app:dirac}

The relevant Lagrangian of the Singlet-doublet Dirac DM model is given by
	\begin{eqnarray}
		\mathcal{L} &\supset& \iota \overline{\Psi} \gamma^\mu D_\mu \Psi + \iota \overline{\chi} \gamma^\mu \partial_\mu \chi - M_{\Psi} \overline{\Psi} \Psi -M_{\chi} \overline{\chi} \chi -
		y\overline{\Psi} \tilde{H} \chi+ {\rm H.c.},
		\label{eq:lag}
	\end{eqnarray}
	where $D_\mu=\partial_\mu-g\frac{\tau_i}{2} W^i_\mu-g^\prime\frac{Y}{2} B_\mu$ and $H$ is the SM Higgs doublet.	After the electro-weak symmetry breaking, the quantum fluctuation around the minima is given as
	\begin{eqnarray}
		H=\begin{pmatrix}
			0\\
			\frac{v+h}{\sqrt{2}}
		\end{pmatrix}.
	\end{eqnarray}
	
	Once Higgs gets a vacuum expectation value, $v$, it induces mixing between the singlet and the neutral component of the doublet fermion through $\overline{\Psi} \Tilde{H} \chi $-coupling. Denoting the mass eigenstates as $\chi_0$ and $\chi_1$ with the mixing angle $\sin\theta$, the transformation from the flavor states to physical states can be written as,
	
	\begin{eqnarray}
		\left(\begin{matrix}
			\psi^0 \\ \chi
		\end{matrix}\right)
		=\left(\begin{matrix}
			\cos\theta & -\sin\theta \\
			\sin\theta & \cos\theta
		\end{matrix}\right)
		\left(\begin{matrix}
			\chi_0 \\  \chi_1
		\end{matrix}\right),
	\end{eqnarray}
	where the mixing angle is given by
	\begin{eqnarray}\label{eq:mixang}
		\tan{2\theta}=  \frac{\sqrt2 y v}{M_\Psi-M_{\chi}} .
	\end{eqnarray}
	
	The mass eigenvalues of the physical states are given as
	
	\begin{eqnarray}
		M_{\chi_0}&=&M_{\Psi} \cos ^2 \theta + \frac{yv}{\sqrt{2}}\sin 2\theta + M_\chi \sin ^2 \theta,~~ \nonumber\\
		M_{\chi_1}&=&M_{\Psi} \sin ^2 \theta - \frac{yv}{\sqrt{2}}\sin 2\theta + M_\chi \cos ^2 \theta\equiv M_{\rm DM}
	\end{eqnarray}
	with a mass-splitting $\Delta M=M_{\chi_0}-M_{\chi_1}$, where, $\chi_1$ is dominantly the singlet fermion and $\chi_0$ is the $\psi^0$.
	From Eq. (\ref{eq:lag}), the interaction among the mass eigenstates ($\chi_0$, $\psi^\pm$ and $\chi_1$), can be expressed as,
	\begin{eqnarray}
		\mathcal{L}_{\rm int} &=& \left(\frac{e_0}{2 \sin\theta_W \cos\theta_W}\right) \left[\sin^2\theta \overline{\chi_1}\gamma^{\mu}Z_{\mu}\chi_1+\cos^2\theta \overline{\chi_0}\gamma^{\mu}Z_{\mu}\chi_0 \right.\left.+\sin\theta \cos\theta(\overline{\chi_1}\gamma^{\mu}Z_{\mu}\chi_0+\overline{\chi_0}\gamma^{\mu}Z_{\mu}\chi_1)\right]   \nonumber \\
		&{}&+\frac{e_0}{\sqrt2\sin\theta_W}\sin\theta \overline{\chi_1}\gamma^\mu W_\mu^+ \psi^- +\frac{e_0}{\sqrt2 \sin\theta_W} \cos\theta \overline{\chi_0}\gamma^\mu W_\mu^+ \psi^-    +\frac{e_0}{\sqrt2 \sin\theta_W} \sin\theta{\psi^+}\gamma^\mu W_\mu^- \chi_1 \nonumber \\
		&{}&+ \frac{e_0}{\sqrt2\sin\theta_W}\cos\theta {\chi^+}\gamma^\mu W_\mu^- \chi_0 - \left(\frac{e_0}{2 \sin\theta_W\cos\theta_W}\right) \cos2\theta_W {\psi^+}\gamma^{\mu}Z_{\mu}\psi^- - e_0 {\psi^+}\gamma^{\mu}A_{\mu}\psi^- \nonumber \\
		&{}& -\frac{y}{\sqrt2}h\left[\sin2\theta(\overline{\chi_1}\chi_1-\overline{\chi_0}\chi_0)+\cos2\theta(\overline{\chi_1}\chi_0+\overline{\chi_0}\chi_1)\right],
	\end{eqnarray}
	
	where $e_0=0.313$ is the electromagnetic coupling constant, and $\theta_W$ is the Weinberg angle.

\section{Lepton number to baryon number conversion}

The asymmetry in the equilibrium number densities of particle $n_i$ over antiparticle $\bar{n}_i$ can be written in the limit $\mu_i/T\ll1$ as,\cite{Harvey:1990qw}
\begin{eqnarray}
    n_i-\bar{n}_i&=&\frac{1}{6}g_iT^3\left(\frac{\mu_i}{T}\right),~~\text{fermion} \nonumber\\
    &=&\frac{1}{3}g_iT^3\left(\frac{\mu_i}{T}\right),~~\text{boson}.
\end{eqnarray}
For a general derivation, we assume that SM consists of $N$ generations of quarks and leptons, $m$ complex Higgs doublets. The chemical potential of the SM fields are assigned as follows: $\mu_W$ for $W^-$, $\mu_0$ for $m$ $H^0$ Higgs fields, $\mu_-$ for $m$ $H^-$ Higgs fields, $\mu_{uL}$ for all left-handed up quark fields, $\mu_{dL}$ for all left-handed down quark fields, $\mu_{uR}$ for all the right-handed up-quark fields, $\mu_{dR}$ for all the right-handed down-quarks fields, $\mu_i$ for the right-handed neutrino fields, $\mu_{iL}$ for the left-handed charged leptons, and $\mu_{iR}$ for all the right-handed charged lepton fields. The details of the BSM fields for both cases are given below.

\subsection{Majorana Case}\label{app:ltob_maj}
We extended the SM by $N'$ generations of vector-like doublet fermions ($\Psi_i$) and $N'$ Majorana right-handed neutrinos ($N_i$) and a singlet real scalar ($\phi$). The chemical potentials for the BSM fields are assigned as: $\mu_{\psi^0}$ for each $\psi^0$, $\mu_{\psi^-}$ for each $\psi^-$, $\mu_{N}$ for all $N_i$, and finally $\mu_\phi$ for $\phi$.\\
Now, the rapid interactions in the early universe enforce the following equilibrium relations among the chemical potentials:
\begin{eqnarray}
\label{eq:Majchemicalreln}
W^- \leftrightarrow H^- + H^0
&\Rightarrow&
\mu_W = \mu_- + \mu_0 \nonumber\\
W^- \leftrightarrow \bar{u}_L + d_L
&\Rightarrow&
\mu_{d_L} = \mu_W + \mu_{u_L} \nonumber\\
W^- \leftrightarrow \bar{\nu}_{iL} + e_{iL}
&\Rightarrow&
\mu_{iL} = \mu_W + \mu_i \nonumber\\
H^0 \leftrightarrow \bar{d}_R + d_L
&\Rightarrow&
\mu_{d_R} = -\mu_0 + \mu_W + \mu_{u_L} \nonumber\\
H^0 \leftrightarrow e_{iL} + \bar{e}_{iR}
&\Rightarrow&
\mu_{iR} = -\mu_0 + \mu_W + \mu_i \nonumber\\
H^0 \leftrightarrow \bar{u}_L + u_R
&\Rightarrow&
\mu_{u_R} = \mu_0 + \mu_{u_L} \\
\psi^-\leftrightarrow W^-+\psi^0&\Rightarrow&\mu_{\psi^-}=\mu_W+\mu_{\psi^0}\nonumber\\
N_i\leftrightarrow H^0+\psi^0&\Rightarrow& \mu_N=\mu_0+\mu_{\psi^0}\nonumber\\
\psi^-\leftrightarrow H^-+N_i&\Rightarrow& \mu_{\psi^-}=\mu_-+\mu_{N}\nonumber\\
\nu_{iL}\leftrightarrow \phi + \psi^0 &\Rightarrow& \mu_i =\mu_{\phi}+\mu_{\psi^0}\nonumber\\
e_{iL}\leftrightarrow \psi^-+\phi&\Rightarrow&\mu_{iL}=\mu_{\psi^-}+\mu_{\phi}\nonumber\\
\phi+\phi\leftrightarrow H^0 + H^{0^*}&\Rightarrow&2\mu_{\phi}=\mu_0-\mu_0\Rightarrow0\nonumber
\end{eqnarray}
Now, the electro-weak $B+L$ anomaly implies the existence of processes that correspond to the creation of $u_Ld_Ld_L\nu_L$ from each generation out of the vacuum. As long as these interactions are rapid, we have
\begin{eqnarray}
N(\mu_{u_L}+2\mu_{d_L})+\sum_i\mu_i
&\Rightarrow&
3N\mu_{u_L}+2N\mu_W+\mu=0.
\end{eqnarray}
Also, since $N_i$ are Majorana RHNs, their chemical potential is zero. Using this fact and the relations from Eq.~\ref{eq:Majchemicalreln}, we find that, $\mu_{\psi^-}=\mu_W-\mu_0$. Let us now express the baryon $(B)$, lepton $(L)$, charge $(Q)$, and third component of weak isospin $(Q_3)$ number densities as
\begin{eqnarray}
B
&=&
3N\left(\frac{1}{3}\mu_{u_L}+\frac{1}{3}\mu_{u_R}\right)
+3N\left(\frac{1}{3}\mu_{d_L}+\frac{1}{3}\mu_{d_R}\right)
\nonumber\\
&=&
4N\mu_{u_L}+2N\mu_W
\end{eqnarray}
\begin{eqnarray}
L
&=&
\sum_i(\mu_i+\mu_{iL}+\mu_{iR})
\nonumber\\
&=&
3\mu+2N\mu_W-N\mu_0
\end{eqnarray}
\begin{eqnarray}
Q
&=&
3N\left(
\frac{2}{3}\mu_{u_L}
-\frac{1}{3}\mu_{d_L}
+\frac{2}{3}\mu_{u_R}
-\frac{1}{3}\mu_{d_R}
\right)
-\left(
\sum_i\mu_{iL}
+
\sum_i\mu_{iR}
\right)
-4\mu_W
-2m\mu_- -N'\mu_{\psi^-}
\nonumber\\
&=&
2N\mu_{u_L}
-2\mu
-(4N+2m+4+N')\mu_W
+(4N+2m+N')\mu_0
\end{eqnarray}
\begin{eqnarray}
Q_3
&=&
3N\left(\frac{1}{2}\mu_{u_L}-\frac{1}{2}\mu_{d_L}\right)
+\sum_i\left(\frac{1}{2}\mu_i-\frac{1}{2}\mu_{iL}\right)
-2*2\mu_W\nonumber\\
&&+2m\left(\frac{1}{2}\mu_+-\frac{1}{2}\mu_0\right) - N'*\frac{1}{2}(\mu_{\psi^-}-\mu_{\psi^0})
\nonumber\\
&=&
-(2N+m+4+\frac{N'}{2})\mu_W
\end{eqnarray}
Now above the critical temperature both $Q$ and $Q_3$ are zero, which will give the $B$ and $L$ in terms of $\mu_{u_L}$ as
\begin{eqnarray}
L
=
\mu_{u_L}N
\left(\frac{8N}{4N+2m+N'}-9
\right)
\end{eqnarray}
\begin{eqnarray}
B &=& 4N\mu_{u_L}
\end{eqnarray}
\begin{eqnarray}
B=
\frac{4}
{\left(\frac{8N}{4N+2m+N'}-9\right)}
L=-\frac{4(4N+2m+N')}{28N+18m+9N'}L
\end{eqnarray}
Taking, $N=3$, $N'=3$ and $m=1$, we have:
\begin{eqnarray}
    B=-0.527 L.
\end{eqnarray}

\subsection{Dirac Case}\label{app:ltob_dir}
We extended the SM by a vector-like doublet fermion ($\Psi_i$) and a vector-like singlet fermion ($\chi$), three singlet complex scalar ($\phi_i$) and $N'$ generations of Dirac right-handed neutrino ($\nu_{R_i}$). The chemical potential for the BSM fields are assigned as: $\mu_{\psi^0}$ for $\psi^0$, $\mu_{\psi^-}$ for $\psi^-$, $\mu_{\chi}$ for $\chi$ and finally $\mu_{\nu R}$ for each $\nu_{R_i}$. Note that, since $\phi_i$ have a soft-breaking term, the chemical potential for $\phi_i$ $\mu_{\phi_i}\rightarrow0$. \\
Now, the rapid interactions in the early universe enforce the following equilibrium relations among the chemical potentials:
\begin{eqnarray}
\label{eq:Diracchemicalreln}
W^- \leftrightarrow H^- + H^0
&\Rightarrow&
\mu_W = \mu_- + \mu_0 \nonumber\\
W^- \leftrightarrow \bar{u}_L + d_L
&\Rightarrow&
\mu_{d_L} = \mu_W + \mu_{u_L} \nonumber\\
W^- \leftrightarrow \bar{\nu}_{iL} + e_{iL}
&\Rightarrow&
\mu_{iL} = \mu_W + \mu_i \nonumber\\
H^0 \leftrightarrow \bar{d}_R + d_L
&\Rightarrow&
\mu_{d_R} = -\mu_0 + \mu_W + \mu_{u_L} \nonumber\\
H^0 \leftrightarrow e_{iL} + \bar{e}_{iR}
&\Rightarrow&
\mu_{iR} = -\mu_0 + \mu_W + \mu_i \nonumber\\
H^0 \leftrightarrow \bar{u}_L + u_R
&\Rightarrow&
\mu_{u_R} = \mu_0 + \mu_{u_L} \\
\psi^-\leftrightarrow W^-+\psi^0&\Rightarrow&\mu_{\psi^-}=\mu_W+\mu_{\psi^0}\nonumber\\
\nu_i\leftrightarrow\phi_i\psi^0 &\Rightarrow& \mu_i=\mu_{\phi_i}+\mu_{\psi^0}\nonumber\\
e_{iL}\leftrightarrow\phi+\psi^-&\Rightarrow& \mu_{iL}=\mu_{\phi_i}+\mu_{\psi^-}\nonumber\\
\chi\leftrightarrow\psi^0+H^0&\Rightarrow&\mu_{\chi}=\mu_{\psi^0}+\mu_0\nonumber\\
\psi^-\leftrightarrow H^-+\chi&\Rightarrow&\mu_{\psi^-}=\mu_-+\mu_{\chi}\nonumber\\
\chi\leftrightarrow\phi+\nu_{R_i}&\Rightarrow& \mu_{\chi}=\mu_{\phi}+\mu_{\nu R}\nonumber
\end{eqnarray}
Now, the electro-weak $B+L$ anomaly implies the existence of processes that correspond to the creation of $u_Ld_Ld_L\nu_L$ from each generation out of the vacuum. As long as these interactions are rapid, we have
\begin{eqnarray}
N(\mu_{u_L}+2\mu_{d_L})+\sum_i\mu_i
&\Rightarrow&
3N\mu_{u_L}+2N\mu_W+\mu=0.
\end{eqnarray}
Using relations from Eq.~\ref{eq:Diracchemicalreln}, we can write $\mu_{\nu R}=\mu_0+\mu_i\implies\mu_{\nu R}=(N\mu_0+\mu)/N$. Let us now express the baryon $(B)$, lepton $(L)$, charge $(Q)$, and third component of weak isospin $(Q_3)$ number densities as
\begin{eqnarray}
B
&=&
3N\left(\frac{1}{3}\mu_{u_L}+\frac{1}{3}\mu_{u_R}\right)
+3N\left(\frac{1}{3}\mu_{d_L}+\frac{1}{3}\mu_{d_R}\right)
\nonumber\\
&=&
4N\mu_{u_L}+2N\mu_W
\end{eqnarray}
\begin{eqnarray}
L
&=&
\sum_i(\mu_i+\mu_{iL}+\mu_{iR})+N'\mu_{\nu R}
\nonumber\\
&=&
\left(3+\frac{N'}{N}\right)\mu+2N\mu_W-\left(N'-N\right)\mu_0
\end{eqnarray}
\begin{eqnarray}
Q
&=&
3N\left(
\frac{2}{3}\mu_{u_L}
-\frac{1}{3}\mu_{d_L}
+\frac{2}{3}\mu_{u_R}
-\frac{1}{3}\mu_{d_R}
\right)
-\left(
\sum_i\mu_{iL}
+
\sum_i\mu_{iR}
\right)
-4\mu_W
-2m\mu_- -\mu_{\psi^-}
\nonumber\\
&=&
2N\mu_{u_L}
-\left(2+\frac{1}{N}\right)\mu
-(4N+2m+5)\mu_W
+(4N+2m)\mu_0
\end{eqnarray}
\begin{eqnarray}
Q_3
&=&
3N\left(\frac{1}{2}\mu_{u_L}-\frac{1}{2}\mu_{d_L}\right)
+\sum_i\left(\frac{1}{2}\mu_i-\frac{1}{2}\mu_{iL}\right)
-2*2\mu_W\nonumber\\
&&+2m\left(\frac{1}{2}\mu_+-\frac{1}{2}\mu_0\right) - \frac{1}{2}(\mu_{\psi^-}-\mu_{\psi^0})
\nonumber\\
&=&
-(2N+m+\frac{9}{2})\mu_W
\end{eqnarray}
Now above the critical temperature both $Q$ and $Q_3$ are zero, which will give the $B$ and $L$ in terms of $\mu_{u_L}$ as
\begin{eqnarray}
L
=
\mu_{u_L}
\left(\left(\frac{8N+3}{4N+2m}\right)(N-N')-3(3N+N')
\right)
\end{eqnarray}
\begin{eqnarray}
B &=& 4N\mu_{u_L}
\end{eqnarray}
\begin{eqnarray}
B=
\frac{4N}{\left(\frac{(8N+3)(N-N')}{4N+2m}\right)-3(3N+N')}L
\end{eqnarray}
Finally, taking, $N=3$, $N'=3$ and $m=1$, we have:
\begin{eqnarray}
    B=-0.33 L.
\end{eqnarray}
%

\providecommand{\href}[2]{#2}\begingroup\raggedright\endgroup

\end{document}